 \documentclass[12pt,preprint]{aastex}

\newcommand{\rs}{r_{_{\rm S}}}

\newcommand{\gapprox}{\lower.4ex\hbox{$\;\buildrel
>\over{\scriptstyle\sim}\;$}}
\newcommand{\lapprox}{\lower.4ex\hbox{$\;\buildrel
<\over{\scriptstyle\sim}\;$}}
\newcommand{\begeq}{\begin{equation}}
\newcommand{\fineq}{\end{equation}}
\newcommand{\begfig}{\begin{figure}}
\newcommand{\finfig}{\end{figure}}
\newcommand{\begeqarray}{\begin{eqnarray}}
\newcommand{\fineqarray}{\end{eqnarray}}
\def\N0{\dot N_0}
\def\Nesc{\dot N_{\rm esc}}
\def\Lesc{L_{\rm esc}}
\newcommand{\Eesc}{E_{\rm esc}}
\newcommand{\NI}{\dot N_{\rm I}}
\newcommand{\NII}{\dot N_{\rm II}}

\newcommand{\CI}{C_{\rm I}}
\newcommand{\CII}{C_{\rm II}}
\newcommand{\QI}{Q_{\rm I}}
\newcommand{\QII}{Q_{\rm II}}

\newcommand{\green}{f_{_{\rm G}}}
\newcommand{\Ljet}{L_{\rm jet}}
\newcommand{\Lshock}{L_{\rm shock}}

\newcommand{\SgrA}{Sgr~A$^*\,$}
\def\pseudophi{\Phi}
\def\sig{\sigma_{_{\rm T}}}

\def\ellprime0{\ell'_0}

\slugcomment{accepted for publication in the Astrophysical Journal}

\shorttitle{Outflows from Advection-Dominated Disks}

\shortauthors{Le \& Becker}

\begin{document}

\title{Particle Acceleration and the Production of Relativistic Outflows
in Advection-Dominated Accretion Disks with Shocks}

\author{Truong Le\altaffilmark{1}}

\affil{E. O. Hulburt Center for Space Research, Naval Research
Laboratory, Washington, DC 20375, USA}

\and

\author{Peter A. Becker\altaffilmark{2}$^,$\altaffilmark{3}}

\affil{Center for Earth Observing and Space Research, George Mason
University, Fairfax, VA 22030-4444, USA}

\altaffiltext{1}{truong.le@nrl.navy.mil}
\altaffiltext{2}{pbecker@gmu.edu}
\altaffiltext{3}{also Department of Physics and Astronomy,
George Mason University, Fairfax, VA 22030-4444, USA}

\begin{abstract}

Relativistic outflows (jets) of matter are commonly observed from
systems containing black holes. The strongest outflows occur in the
radio-loud systems, in which the accretion disk is likely to have an
advection-dominated structure. In these systems, it is clear that the
binding energy of the accreting gas is emitted primarily in the form of
particles rather than radiation. However, no comprehensive model for the
disk structure and the associated outflows has yet been produced. In
particular, none of the existing models establishes a direct physical
connection between the presence of the outflows and the action of a
microphysical acceleration mechanism operating in the disk. In this
paper we explore the possibility that the relativistic protons powering
the jet are accelerated at a standing, centrifugally-supported shock
in the underlying accretion disk via the first-order Fermi mechanism.
The theoretical analysis employed here parallels the early studies of
cosmic-ray acceleration in supernova shock waves, and the particle
acceleration and disk structure are treated in a coupled,
self-consistent manner based on a rigorous mathematical approach.
We find that first-order Fermi acceleration at standing shocks in
advection-dominated disks proves to be a very efficient means for
accelerating the jet particles. Using physical parameters appropriate
for M87 and \SgrA, we verify that the jet kinetic luminosities computed
using our model agree with estimates based on observations of the
sources.

\end{abstract}

% The different journals have different requirements for keywords. The
% keywords.apj file, found on aas.org in the pubs/aastex-misc directory,
% contains a list of keywords used with the ApJ and Letters.  These are
% usually assigned by the editor, but authors may include them in their
% manuscripts if they wish.


\section{INTRODUCTION}

A large body of observational evidence has established that
extragalactic relativistic jets are commonly associated with radio-loud
active galactic nuclei (AGNs), which may contain hot,
advection-dominated accretion disks. However, the precise nature of the
mechanism responsible for transferring the gravitational potential
energy from the infalling matter to the small population of nonthermal
particles that escape to form the jet is not yet clear (see, e.g., Livio
1999). The most promising jet acceleration scenarios proposed so far are
the Blandford-Znajek mechanism (Blandford \& Znajek 1977) and the
electromagnetic cocoon model (Lovelace 1976; Blandford \& Payne 1982),
both of which involve the extraction of energy from the rotation of the
black hole in order to power the outflow. While conceptually attractive,
one finds that the complex physics involved in these models tends to
obscure the nature of the fundamental microphysical processes. In
particular, the introduction of the relativistic particles that escape
to form the jet is usually made in an ad hoc manner without any
reference to the dynamics of the associated accretion disk, although
recent magnetohydrodynamical simulations carried out by De Villiers et
al. (2005) and McKinney \& Gammie (2004) have achieved a higher level of
self-consistency. Given the relative complexity of the electromagnetic
models, it is natural to ask whether the outflows can be explained in
terms of well-understood microphysical processes operating in the hot,
tenuous disk, such as the possible acceleration of the jet particles at
a standing accretion shock.

It has been known for some time that inviscid accretion disks can
display both shocked and shock-free (i.e., smooth) solutions depending
on the values of the energy and angular momentum per unit mass in the
gas supplied at a large radius (e.g., Chakrabarti 1989a; Chakrabarti \&
Molteni 1993; Kafatos \& Yang 1994; Lu \& Yuan 1997; Das, Chattopadhyay,
\& Chakrabarti 2001). Shocks can also exist in viscous disks if the
viscosity is relatively low (Chakrabarti 1996; Lu, Gu, \& Yuan 1999),
although smooth solutions are always possible for the same set of
upstream parameters (Narayan, Kato, \& Honma 1997; Chen, Abramowicz, \&
Lasota 1997). Hawley, Smarr, \& Wilson (1984a, 1984b) have shown through
general relativistic simulations that if the gas is falling with some
rotation, then the centrifugal force can act as a ``wall,'' triggering
the formation of a shock. Furthermore, the possibility that shock
instabilities may generate the quasi-periodic oscillations (QPOs)
observed in some sources containing black holes has been pointed out by
Chakrabarti, Acharyya, \& Molteni (2004), Lanzafame, Molteni, \&
Chakrabarti (1998), Molteni, Sponholz, \& Chakrabarti (1996), and
Chakrabarti \& Molteni (1995). Nevertheless, shocks are ``optional''
even when they are allowed, and one is always free to construct models
that avoid them. However, in general the shock solution possesses a
higher entropy content than the shock-free solution, and therefore we
argue based on the second law of thermodynamics that when possible, the
shocked solution represents the preferred mode of accretion (Becker \&
Kazanas 2001; Chakrabarti \& Molteni 1993).

Our primary objective in this paper is to explore the consequences of
the presence of a shock in an ADAF disk for the acceleration of the
nonthermal particles in the observed jets. The question of whether or
not viscosity needs to be included in the disk model is difficult to
answer in general. Several authors have shown that shock solutions are
possible in viscous (e.g., Chakrabarti 1990, 1996; Lu, Gu, \& Yuan 1999;
Chakrabarti \& Das 2004) as well as inviscid disks (e.g., Chakrabarti
1989a, 1989b; Abramowicz \& Chakrabarti 1990; Yang \& Kafatos 1995;
Chakrabarti 1996; Das, Chattopadhyay, \& Chakrabarti 2001). In
particular, Chakrabarti (1990) and Chakrabarti \& Das (2004)
demonstrated that shocks can exist in viscous disks if the angular
momentum and the viscosity are relatively low. Since the acceleration of
particles in shocked disks has never been investigated before, in this
first study we shall focus on inviscid flows containing isothermal
shocks (e.g., Chakrabarti 1989a; Kafatos \& Yang 1994; Lu \& Yuan 1997),
while deferring the treatment of viscous disks to future work. However,
it is clearly important to address the potential connection between this
idealized, inviscid calculation and the physical properties of real
accretion disks, which undoubtedly have nonzero viscosity. We argue that
the results presented here should be qualitatively similar to those
obtained in a viscous disk provided a shock is present, in which case
efficient first-order Fermi acceleration is expected to occur. While the
possible existence of standing shocks in viscous disks is a
controversial issue at the present time, we believe that the work of
Chakrabarti (1990, 1996), Lu, Gu, \& Yuan (1999), and Chakrabarti \& Das
(2004) provides sufficient support for the possibility to motivate the
present investigation.

Although the effect of a standing shock in {\it heating} the gas in the
post-shock region has been examined by a number of previous authors for
both viscid (Chakrabarti \& Das 2004; Lu, Gu, \& Yuan 1999; Chakrabarti
1990) and inviscid (e.g., Lu \& Yuan 1997, 1998; Yang \& Kafatos 1995;
Abramowicz \& Chakrabarti 1990) disks, the implications of the shock for
the acceleration of {\it nonthermal} particles in the disk have not been
considered in detail before. However, a great deal of attention has been
focused on particle acceleration in the vicinity of supernova-driven
shock waves as a possible explanation for the observed cosmic-ray energy
spectrum (Blandford \& Ostriker 1978; Jones \& Ellison 1991). In the
present paper we consider the analogous process occurring in hot,
advection-dominated accretion flows (ADAFs) around black holes. These
disks are ideal sites for first-order Fermi acceleration at shocks
because the plasma is collisionless and therefore a small fraction of
the particles can gain a great deal of energy by repeatedly crossing the
shock. Shock acceleration in the disk therefore provides an intriguing
possible explanation for the powerful outflows of relativistic particles
observed in many radio-loud systems (Le \& Becker 2004).

The dynamical model for the disk/shock/outflow employed here was
discussed by Le \& Becker (2004), who demonstrated that the predicted
kinetic power in the jets agrees with the observational estimates for
M87 and \SgrA. Here we present a more detailed development of the
dynamical model, including a careful examination of the implications of
the shock acceleration process for the evolution of the relativistic
particle distribution in the disk and the jet. The number and energy
densities of the relativistic particles are determined along with the
hydrodynamical structure of the disk in a self-consistent manner by
solving the fluid dynamical conservation equations and the transport
equation simultaneously using a rigorous mathematical approach. In this
sense, the model presented here represents a new type of synthesis
between studies of accretion dynamics and particle transport.

The remainder of the paper is organized as follows. In \S~2 we
discuss the ADAF model assumptions and the possibility of shock
acceleration in ADAF disks, and the general structure of the
disk/shock model is examined in \S~3. The isothermal shock jump
conditions and the asymptotic variations of the physical
parameters at both large and small radii are discussed in \S~4. In
\S~5 we analyze the steady-state transport equation governing the
distribution of the relativistic particles in the disk and the
jet. Solutions for the number and energy density distributions of
the relativistic particles are obtained in \S~6, and detailed
applications to the disks/outflows in M87 and \SgrA are presented
in \S~7. The astrophysical implications of our results are
discussed in \S~8.

\section{MODEL BACKGROUND}

Accretion onto a black hole involves differentially-rotating flows
in which the viscosity plays an essential role in transporting angular
momentum outward, thereby allowing the accreting gas to spiral in toward
the central mass (Pringle 1981). In the ADAF model, it is assumed
that the mass accretion rate is much smaller than the Eddington rate,
\begeq
\dot M_{\rm E} \equiv  c^{-2} \, \beta^{-1} \, L_{\rm E}
= 2.2 \times 10^{-9} \, \beta^{-1} \left({M \over M_\odot}
\right) \ \ {\rm M_\odot \ yr^{-1}}
\ ,
\label{eqnew1}
\fineq
where the efficiency parameter $\beta$ is of order $\sim 10\%$, and the
Eddington luminosity is defined by $L_{\rm E} \equiv 4 \pi \, G M m_p \,
c/\sig$ for pure, fully-ionized hydrogen, with $\sig$, $M$, $m_p$, and
$c$ denoting the Thomson cross section, the black-hole mass, the proton
mass, and the speed of light, respectively. Due to the sub-Eddington
accretion rates in these systems, the plasma is rather tenuous, and this
strongly inhibits the efficiency of two-body radiative processes such as
free-free emission. The gas is therefore unable to cool effectively
within an accretion time, and consequently the gravitational potential
energy dissipated by viscosity is stored in the gas as thermal energy
instead of being radiated away (e.g., Narayan, Kato, \& Honma 1997). The
low density also reduces the level of Coulomb coupling between the ions
and the electrons, resulting in a two-temperature configuration with the
ion temperature ($T_i \sim 10^{12} \rm K$) close to the virial value,
and a much lower electron temperature ($T_e \sim 10^9 \rm K$). In this
scenario, most of the energy is advected across the horizon into the
black hole, and the resulting X-ray luminosity is far below the
Eddington value (Becker \& Le 2003; Becker \& Subramanian 2005).

When the ion temperature is close to the virial temperature, as in
ADAFs, the disk is gravitationally unbound (e.g., Narayan, Kato, \&
Honma 1997; Blandford \& Begelman 1999; Becker, Subramanian, \& Kazanas
2001). It follows that the original ADAF model was not entirely
self-consistent since it neglected outflows. This motivated Blandford \&
Begelman (1999) to propose the self-similar advection-dominated
inflow-outflow solution (ADIOS) to address the question of
self-consistency by including the possibility of powerful winds that
carry away mass, energy, and angular momentum. In this Newtonian,
nonrelativistic model, the dynamical solutions are not applicable near
the event horizon, and therefore the ADIOS approach cannot be used to
obtain a global understanding of the disk structure. This led Becker,
Subramanian, \& Kazanas (2001) to modify the ADIOS scenario to include
general relativistic effects by replacing the Newtonian potential with
the pseudo-Newtonian form (Paczy\'nski \& Wiita 1980)
\begeq
\pseudophi(r) \equiv {-GM \over r-\rs}
\ ,
\label{eqnew2}
\fineq
where $\rs=2GM/c^2$ is the Schwarzschild radius for a black hole
of mass $M$. This modified model is known as the self-similar
relativistic advection-dominated inflow-outflow solution (RADIOS).
Despite the success of the self-similar RADIOS model in describing
the general features of the disk/outflow structure, it does not
provide a comprehensive picture since no explicit microphysical
acceleration mechanism is included. It is therefore natural to
explore possible extensions to the ADAF scenario that incorporate
a concrete acceleration mechanism capable of powering the
outflows.

%\subsection{Importance of Collisions and Viscosity}

%\subsection{Shock Acceleration in Accretion Disks}

The idea of shock acceleration in the environment of AGNs was first
suggested by Blandford \& Ostriker (1978). Subsequently, Protheroe \&
Kazanas (1983) and Kazanas \& Ellison (1986) investigated shocks in
spherically-symmetric accretion flows as a possible explanation for the
energetic radiation emitted by many AGNs. However, in these papers the
acceleration of the particles was studied without the benefit of a
detailed transport equation, and the assumption of spherical symmetry
precludes the treatment of acceleration in disks. The state of the
theory was advanced by Webb \& Bogdan (1987) and Spruit (1987), who
employed a transport equation to solve for the distribution of energetic
particles in a spherical accretion flow characterized by a self-similar
velocity profile terminating at a standing shock. While more
quantitative in nature than the earlier models, these solutions are not
applicable to disks since the geometry is spherical and the velocity
distribution is inappropriate. Hence none of these previous models can
be used to develop a single, global, self-consistent picture for the
acceleration of relativistic particles in an accretion disk containing a
shock.

The success of the diffusive (first-order Fermi) shock acceleration
model in the cosmic-ray context suggests that the same mechanism may be
responsible for powering the outflows commonly observed in radio-loud
systems containing black holes. As a preliminary step in evaluating the
potential relevance of shock acceleration as a possible explanation for
the observed outflows, we need to consider the basic physical properties
of the hot plasma in ADAF disks. One of the critical issues for
determining the efficiency of shock acceleration in accretion disks is
the role of particle-particle collisions in thermalizing the high-energy
ions. The mean free path for ion-ion collisions is given in cgs units by
(Subramanian, Becker, \& Kafatos 1996)
\begeq
\lambda_{ii} = 1.8 \times 10^5 \, {T_i^2 \over N_i \ln \Lambda}
\ ,
\label{eqnew3}
\fineq
where $N_i$ and $T_i$ denote the thermal ion number density and
temperature, respectively, and $\ln \Lambda$ is the Coulomb logarithm.
In ADAF disks, $\lambda_{ii}$ greatly exceeds the vertical thickness of
the disk, and therefore the shock and the flow in general are
collisionless. However, the mean free path $\lambda_{\rm mag}$ for
collisions between ions and magnetohydrodynamical (MHD) waves is much
shorter than $\lambda_{ii}$ for the thermal particles, and it is much
longer than $\lambda_{ii}$ for the relativistic particles (Ellison \&
Eichler 1984; Subramanian, Becker, \& Kafatos 1996). The increase in
$\lambda_{\rm mag}$ with increasing particle energy reflects the fact
that the high-energy particles will interact only with the
highest-energy MHD waves. The low-energy background particles therefore
tend to thermalize the energy they gain in crossing the shock due to
collisions with magnetic waves. Conversely, the relativistic particles
are able to diffuse back and forth across the shock many times, gaining
a great deal of energy while avoiding thermalization due to the longer
mean free path.

The probability of multiple shock crossings decreases exponentially with
the number of crossings, and the mean energy of the particles increases
exponentially with the number of crossings. This combination of factors
naturally gives rise to a power-law energy distribution, which is a
general characteristic of Fermi processes (Fermi 1954). Two effects
limit the maximum energy that can be achieved by the particles. First,
at very high energies the particles will tend to lose energy to the
waves due to recoil. Second, the mean free path $\lambda_{\rm mag}$ will
eventually exceed the thickness of the disk as the particle energy is
increased, resulting in escape from the disk without further
acceleration.

\begfig[t]
\centering
\includegraphics[scale=0.6]{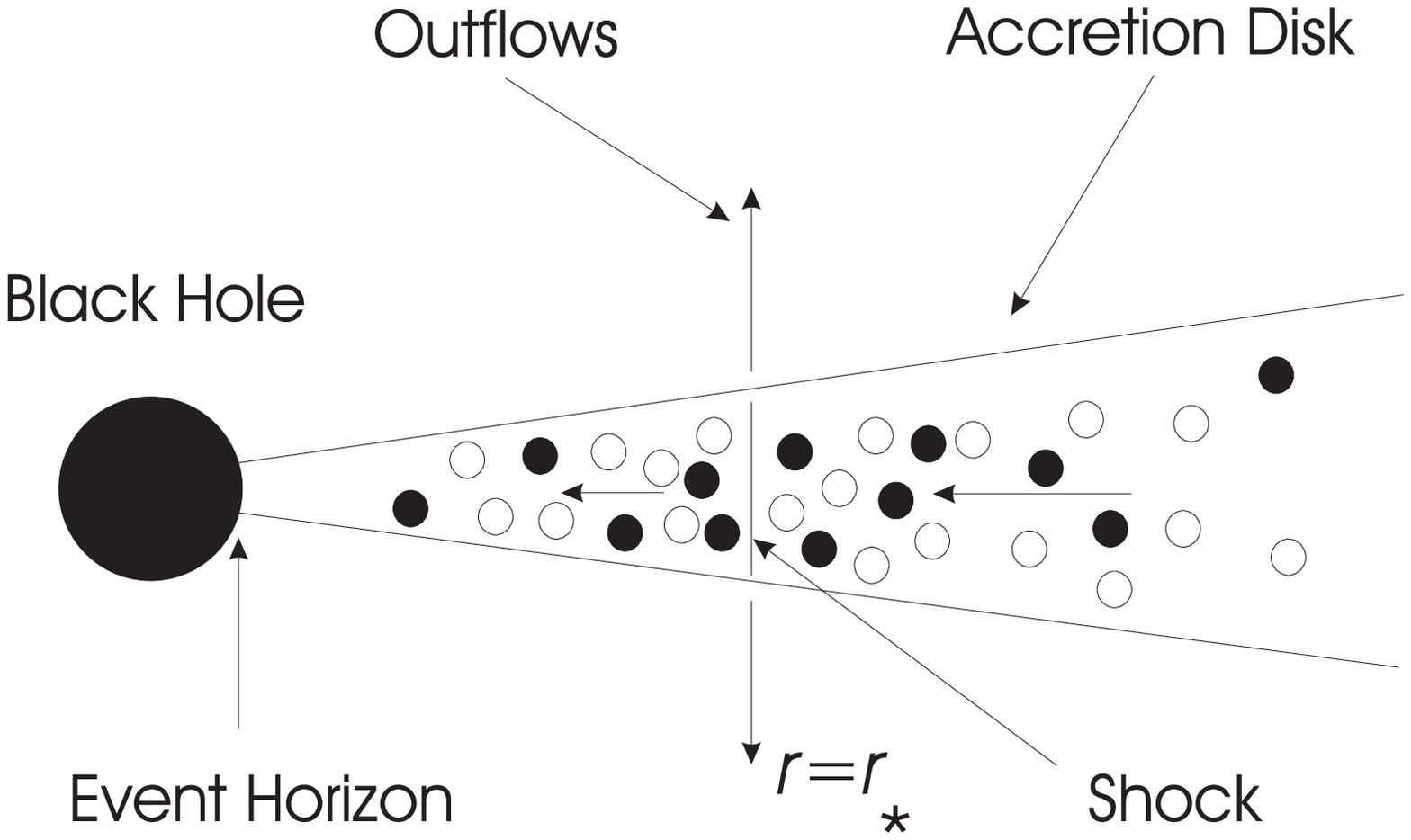}
\caption{Schematic representation of our disk/shock/outflow model. The
filled circles in the disk represent the test particles, and the
unfilled circles represent the background gas. The test particles are
injected at the shock location.}
\label{chpt3_fig2}
\finfig

\section{TRANSONIC FLOW STRUCTURE}

As discussed in \S~1, various authors have established that shocks can
exist in both viscid and inviscid disks. In this first study of particle
acceleration in shocked disks, we shall focus on the inviscid case
because it is the most straightforward to analyze from a mathematical
viewpoint, and also because it serves to illustrate the basic physical
principles involved. Moreover, we expect that the results obtained in
the viscous case will be qualitatively similar to those presented here
since efficient Fermi acceleration will occur whether or not viscosity
is present, provided the flow contains a shock. The equations governing
the disk structure can yield solutions that include three possible types
of standing shocks, namely (i) Rankine-Hugoniot shocks, where the
effective cooling processes are so inefficient that no energy is lost
from the surface of the disk, (ii) isentropic shocks, where the entropy
generated at the shock is comparable to the amount radiated away, and
(iii) isothermal shocks, where the cooling processes are so efficient
that the post-shock sound speed and disk thickness remain the same as
the pre-shock values. In the isothermal case, the shock must radiate
away both energy and entropy through the upper and lower surfaces of the
disk (e.g., Chakrabarti 1989a, 1989b; Abramowitz $\&$ Chakrabarti 1990).
This renders the isothermal shock model particularly useful from the
point of view of modeling outflows, since the energy lost from the shock
can be identified with that powering the jet. On the other hand,
Rankine-Hugoniot shocks cannot be used if we are interested in any kind
of escape. The isentropic shock is an intermediate case. In this paper,
we shall focus exclusively on the isothermal shock model since this case
provides the strongest potential connection with the observed outflows.

The model considered here is depicted schematically in
Figure~\ref{chpt3_fig2}. In this scenario, the gas is accelerated
gravitationally toward the central mass, and experiences a shock
transition due to an obstruction near the event horizon. The obstruction
is provided by the ``centrifugal barrier,'' which is located between the
inner and outer sonic points. Particles from the high-energy tail of the
background Maxwellian distribution are accelerated at the shock
discontinuity via the first-order Fermi mechanism, resulting in the
formation of a nonthermal, relativistic particle distribution in the
disk. The spatial transport of the energetic particles within the disk
is a stochastic process based on a three-dimensional random walk through
the accreting background gas. Consequently, some of the accelerated
particles diffuse to the disk surface and become unbound, escaping
through the upper and lower edges of the cylindrical shock to form the
outflow, while others diffuse outward radially through the disk or
advect across the event horizon into the black hole.

In order to analyze the connection between the disk/shock model and the
transport/acceleration of the relativistic particles, we consider the
set of physical conservation equations employed by Chakrabarti (1989a)
and Abramowicz \& Chakrabarti (1990), who investigated the structure of a
one-dimensional, steady-state, axisymmetric, inviscid accretion flow
based on the vertically-averaged conservation equations. The effects of
general relativity are incorporated in an approximate manner by
utilizing the pseudo-Newtonian form for the gravitational potential per
unit mass given by equation~(\ref{eqnew2}). The use of such a potential
allows one to investigate the complicated physical processes taking
place in the accretion disk within the context of a semi-classical
framework while maintaining good agreement with fully relativistic
calculations (see, e.g., Narayan, Kato, \& Honma 1997; Becker \&
Subramanian 2005). The pseudo-Newtonian potential correctly reproduces
the radius of the event horizon, the marginally bound orbit, and the
marginally stable orbit (Paczy\'nski $\&$ Wiita 1980). Furthermore, the
dynamics of freely-falling particles near the event horizon computed
using this potential agree perfectly with the results obtained using the
Schwarzschild metric, although time dilation is not included
(Becker \& Le 2003).

\subsection{Transport Rates}

Becker \& Le (2003) and Becker \& Subramanian (2005) demonstrated that
three integrals of the flow are conserved in viscous ADAF disks, namely,
the mass transport rate
\begeq
\dot M = 4 \pi r H \rho \, v \ ,
\label{eqnew4}
\fineq
the angular momentum transport rate
\begeq
\dot J = \dot M r^2 \, \Omega - \mathcal{G} \ ,
\label{eqnew5}
\fineq
and the energy transport rate
\begeq
\dot E = - \mathcal{G} \, \Omega +
\dot M\left({1 \over 2} \, v_{\phi}^2 + {1 \over 2} \, v^2
          + {P+U \over \rho} + \pseudophi \right) \ ,
\label{eqnew6}
\fineq
where $\rho$ is the mass density, $v$ is the radial velocity (defined to
be positive for inflow), $\Omega$ is the angular velocity, ${\cal G}$ is
the torque, $H$ is the disk half-thickness, $v_{\phi}=r \, \Omega$ is
the azimuthal velocity, $U$ is the internal energy density, and
$P = (\gamma-1)U$ is the pressure. Each of the various quantities represents
a vertical average over the disk structure. We also assume that the
ratio of specific heats, $\gamma$, maintains a constant value throughout
the flow. Note that all of the transport rates $\dot M$, $\dot J$, and
$\dot E$ are defined to be positive for inflow.

The torque ${\cal G}$ is related to the gradient of the angular velocity
$\Omega$ via the usual expression (e.g., Frank et al. 2002)
\begeq
\mathcal{G} =
- 4 \pi r^3 H \rho \, \nu {d\Omega \over dr} \ ,
\label{eqnew7}
\fineq
where $\nu$ is the kinematic viscosity. The disk half-thickness $H$ is
given by the standard hydrostatic prescription
\begeq
H(r) = {a \over \Omega_{\rm K}} \ ,
\label{eqnew8}
\fineq
where $a$ represents the adiabatic sound speed, defined by
\begeq
a(r) \equiv \left({\gamma P \over \rho}\right)^{1/2}  \ ,
\label{eqnew9}
\fineq
and $\Omega_{\rm K}$ denotes the Keplerian angular velocity of matter
in a circular orbit at radius $r$ in the pseudo-Newtonian potential
(eq.~[\ref{eqnew2}]), defined by
\begeq
\Omega_{\rm K}^2 \equiv {GM \over r(r-\rs)^2}
= {1 \over r} {d\pseudophi \over dr} \ .
\label{eqnew10}
\fineq

The quantities $\dot M$ and $\dot J$ are constant throughout the flow,
and therefore they represent the rates at which mass and angular
momentum, respectively, enter the black hole. The energy transport rate
$\dot E$ generally remains constant, although it will jump at the
location of an isothermal shock if one is present in the disk. We can
eliminate the torque ${\cal G}$ between equations~(\ref{eqnew5}) and
(\ref{eqnew6}) and combine the result with equation~(\ref{eqnew9}) to
express the energy transport per unit mass as
\begeq
\epsilon \equiv {\dot E \over \dot M} = {1 \over 2} \, v^2
- {1 \over 2} \, {\ell^2 \over r^2} +  {\ell_0 \ell \over r^2}
+ {a^2 \over \gamma-1} + \pseudophi \ ,
\label{eqnew11}
\fineq
where $\ell(r) \equiv r^2 \, \Omega(r)$ and $\ell_0 \equiv \dot J/\dot
M$ denote the specific angular momentum at radius $r$ and the (constant)
angular momentum transport per unit mass, respectively. Note that
equation (\ref{eqnew11}) is valid for both viscid and inviscid flows.

\subsection{Inviscid Flow Equations}

In the present application, viscosity is neglected, and therefore
${\cal G} = 0$ and the specific angular momentum is given by
\begeq
\ell(r) = \ell_0 = {\rm constant}
\label{eqnew12}
\fineq
throughout the disk. It follows that the flow is purely adiabatic,
except at the location of a possible isothermal shock (Becker \& Le
2003). In the inviscid case, equation (\ref{eqnew11}) reduces to
\begeq
\epsilon = {1 \over 2} \, v^2 + {1 \over 2} \, {\ell^2
\over r^2} + {a^2 \over \gamma-1} + \pseudophi \ .
\label{eqnew12a}
\fineq
The resulting disk/shock model depends on three free parameters,
namely the energy transport per unit mass $\epsilon$, the specific
heats ratio $\gamma$, and the specific angular momentum $\ell$.
The value of $\epsilon$ will jump at the location of an isothermal
shock if one exists in the disk, but the value of $\ell$ remains
constant throughout the flow. This implies that the specific
angular momentum of the particles escaping through the upper and
lower surfaces of the cylindrical shock must be equal to the
average value of the specific angular momentum for the particles
remaining in the disk, and therefore the outflow exerts no torque
on the disk (Becker, Subramanian, \& Kazanas 2001). Since the flow
is purely adiabatic in the absence of viscosity, the pressure and
density variations are coupled according to
\begeq
P = D_0 \, \rho^\gamma \ ,
\label{eqnew13}
\fineq
where $D_0$ is a parameter related to the specific entropy that remains
constant except at the location of the isothermal shock if one is
present.

By combining equations~(\ref{eqnew4}), (\ref{eqnew8}), (\ref{eqnew9}),
(\ref{eqnew10}), and (\ref{eqnew13}), we find that the quantity
\begeq
K \equiv r^{3/2} \, (r - \rs) \, v \, a^{(\gamma+1)/(\gamma-1)}
\label{eqnew14}
\fineq
is conserved throughout an adiabatic disk, except at the location of an
isothermal shock. Following Becker \& Le (2003), we refer to $K$ as the
``entropy parameter,'' and we note that the entropy per particle $S$ is
related to $K$ via
\begeq
S = k  \ln K + c_0
\ ,
\label{eqnew15}
\fineq
where $k$ is the Boltzmann constant and $c_0$ is a constant that depends
on the composition of the gas but is independent of its state.

\subsection{Critical Conditions}

By combining equations~(\ref{eqnew12a}) and (\ref{eqnew14}), one
can solve for the flow velocity $v$ as a function of $r$ using a
simple root-finding procedure. However, in order to understand the
implications of the transonic (critical) nature of the accretion
flow, we must also analyze the properties of the ``wind
equation,'' which is the first-order differential equation
governing the flow velocity $v$. By differentiating
equation~(\ref{eqnew12a}) with respect to $r$, we obtain the
steady-state radial momentum equation
\begeq
v {dv \over dr} = {\ell^2 \over r^3} - {d\pseudophi \over dr}
- \left({2 \, a \over \gamma-1}\right) {da \over dr} \ .
\label{eqnew16}
\fineq
The derivative of the sound speed appearing on the right-hand side of
this expression can be evaluated by using equations~(\ref{eqnew10}) and
(\ref{eqnew14}) to write
\begeq
{1 \over a} {da \over dr} = \left({\gamma-1 \over
\gamma+1}\right)\left[ {1 \over \Omega_{\rm K}} {d\Omega_{\rm K}
\over dr} - {1 \over v} {dv \over dr} - {1 \over r} \right] \ .
\label{eqnew17}
\fineq
We can now construct the wind equation by combining equations~(\ref{eqnew10}),
(\ref{eqnew16}), and (\ref{eqnew17}), which yields
\begeq
{1 \over v} {dv \over dr} = {N \over D} \ ,
\label{eqnew18}
\fineq
where the numerator and denominator functions $N$ and $D$ are given by
\begeq
N = {G M \over (r-\rs)^2} - {\ell^2 \over r^3}
+ {a^2 \over \gamma+1} \left[{3 \rs - 5 r \over r(r-\rs)}
\right]
\ , \ \ \ \ \ \
D = {2 \, a^2 \over \gamma+1} - v^2 \ .
\label{eqnew19}
\fineq
The simultaneous vanishing of $N$ and $D$ yields the critical conditions
\begeq
{G M \over (r_c-\rs)^2}
- {\ell^2 \over r_c^3}
+ {a_c^2 \over \gamma+1} \left[{3 \rs - 5 r_c \over
r_c(r_c-\rs)} \right] = 0 \ ,
\label{eqnew20}
\fineq
and
\begeq
{2 a_c^2 \over \gamma+1} - v_c^2 = 0 \ ,
\label{eqnew21}
\fineq
where $v_c$ and $a_c$ denote the values of the velocity and the sound
speed at the critical radius, $r=r_c$.

\subsection{Critical Point Analysis}

Equations (\ref{eqnew20}) and (\ref{eqnew21}) can be solved
simultaneously to express $v_c$ and $a_c$ as explicit functions of the
critical radius $r_c$, which yields
\begeq
v^2_c = 2\left[{GM r^3_c-\ell^2(r_c-\rs)^2 \over
(5r_c-3\rs)(r_c-\rs)r^2_c}\right] \ ,
\label{eqnew22}
\fineq
and
\begeq
a^2_c = (\gamma+1)\left[{GM r^3_c-\ell^2(r_c-\rs)^2
\over (5r_c-3\rs)(r_c-\rs)r^2_c}\right] \ .
\label{eqnew23}
\fineq
The corresponding value of the entropy parameter $K$ at the critical
point is given by (see eq.~[\ref{eqnew14}])
\begeq
K_c = r_c^{3/2} \, (r_c - \rs) \, v_c \, a_c^{(\gamma+1)/(\gamma-1)}
\ .
\label{eqnew24}
\fineq
By using equations~(\ref{eqnew22}) and (\ref{eqnew23}) to
substitute for $v$ and $a$ in equation~(\ref{eqnew12a}), we can
express the energy transport parameter $\epsilon$ in terms of
$r_c$, $\ell$, and $\gamma$, obtaining
\begeq
\epsilon = {1 \over 2} \, {\ell^2 \over r_c^2}
- {G M \over r_c - \rs} + {2 \gamma \over \gamma-1}
\left[{GM r_c^3 - \ell^2 (r_c - \rs)^2 \over
r_c^2 (r_c - \rs) (5 r_c - 3 \rs)}\right] \ .
\label{eqnew25}
\fineq
This expression can be rewritten as a quartic equation for $r_c$ of
the form
\begeq
\mathcal{N} \, r^4_c - \mathcal{O} \, r^3_c + \mathcal{P} \, r^2_c
- \mathcal{Q} \, r_c + \mathcal{R} = 0 \ ,
\label{eqnew26}
\fineq
where
\begeqarray
\mathcal{N}& = & 5 \, \epsilon \ , \ \ \ \
\mathcal{O} = 16 \, \epsilon - 3 + {2 \over \gamma-1} \ , \nonumber \\
\mathcal{P} & = & 12 \, \epsilon +
{1 \over 2}\left({5 - \gamma \over \gamma-1}\right)\ell^2 - 6 \ ,
\label{eqnew27} \\
\mathcal{Q} & = & \left({8 \over \gamma-1}\right)\ell^2 \ ,
\nonumber \\
\mathcal{R} & = & \left({2 \gamma + 6 \over \gamma-1}\right)\ell^2 \ ,
\nonumber
\fineqarray
and we have utilized natural gravitational units with $GM=c=1$ and
$\rs=2$. These equations agree with the corresponding results derived by
Das, Chattopadhyay, \& Chakrabarti (2001). The four solutions for $r_c$
in terms of the three fundamental parameters $\epsilon$, $\ell$, and
$\gamma$ can be obtained analytically using the standard formulas for
quartic equations (e.g., Abramowitz \& Stegun 1970). We refer to the
roots using the notation $r_{c1}$, $r_{c2}$, $r_{c3}$, and $r_{c4}$ in
order of decreasing radius.

The critical radius $r_{c4}$ always lies inside the event horizon and is
therefore not physically relevant, but the other three are located
outside the horizon. The type of each critical point is determined by
computing the two possible values for the derivative $dv/dr$ at the
corresponding location using L'H\^opital's rule and then checking to see
whether they are real or complex. We find that both values are complex
at $r_{c2}$, and therefore this is an O-type critical point, which does
not yield a physically acceptable solution. The remaining roots $r_{c1}$
and $r_{c3}$ each possess real derivatives, and are therefore physically
acceptable sonic points, although the types of accretion flows that can
pass through them are different. Specifically, $r_{c3}$ is an X-type
critical point, and therefore a smooth, global, shock-free solution
always exists in which the flow is transonic at $r_{c3}$ and then
remains supersonic all the way to the event horizon. On the other hand,
$r_{c1}$ is an $\alpha$-type critical point, and therefore any accretion
flow originating at a large distance that passes through this point must
display a shock transition below $r_{c1}$ (Abramowicz \& Chakrabarti
1990). After crossing the shock, the subsonic gas must pass through
another ($\alpha$-type) critical point before it enters the black hole
since the flow has to be supersonic at the event horizon (Weinberg
1972).

\subsection{Shock-Free Solutions}

Even when a shock can exist in the flow, it is always possible to
construct a globally smooth (shock-free) solution using the same
set of parameters. Smooth flows must pass through the inner
critical point located at radius $r_{c3}$, which is calculated
using the quartic equation~(\ref{eqnew26}) for given values of
$\epsilon$, $\ell$, and $\gamma$. The corresponding values for the
critical velocity, $v_{c3}$, the critical sound speed, $a_{c3}$,
and the critical entropy, $K_{c3}$, are computed using
equations~(\ref{eqnew22}), (\ref{eqnew23}), and (\ref{eqnew24}),
respectively. Since the flows treated here are inviscid, they have
a conserved value for the entropy parameter $K$
(eq.~[\ref{eqnew14}]) unless a shock is present. Hence in a
smooth, shock-free flow, the value of $K$ is everywhere equal to
the critical value $K_{c3}$. The structure of the velocity profile
in a shock-free disk can therefore be determined using a simple
root-finding procedure as follows. By eliminating the sound speed
$a$ between equations~(\ref{eqnew12a}) and (\ref{eqnew14}), we
obtain the equivalent expression
\begeq
\epsilon = {1 \over 2} \, v^2 + {1 \over 2} \,
{\ell^2 \over r^2} + {1 \over \gamma-1} \,
\left[{K_{c3}^2 \over r^3 (r-\rs)^2 \, v^2}\right]^{(\gamma-1)/(\gamma+1)}
+ \pseudophi \ ,
\label{eqnew28}
\fineq
where we have set $K=K_{c3}$. In general, at any radius $r$,
equation~(\ref{eqnew28}) yields one subsonic root and one supersonic
root for the velocity. The subsonic solution is chosen for $r > r_{c3}$,
and the supersonic solution is selected for $r < r_{c3}$. Once the
velocity profile $v(r)$ has been computed, we can obtain the
corresponding sound speed distribution $a(r)$ by utilizing
equation~(\ref{eqnew14}) with $K=K_{c3}$. Note that the velocity and sound
speed solutions can also be calculated by integrating the wind
equation~(\ref{eqnew18}) numerically, and the results obtained using this
approach agree with the root-finding method.

\section{ISOTHERMAL SHOCK MODEL}

Our primary goal in this paper is to analyze the acceleration of
relativistic particles due to the presence of a standing, isothermal
shock in an accretion disk. Hence we are interested in flows that pass
smoothly through the outer critical radius $r_{c1}$, and then experience
a velocity discontinuity at the shock location, which we refer to as
$r_*$. In order to form self-consistent global models, we will first
need to understand how the structure of the disk responds to the
presence of a shock. This requires analysis of the shock jump
conditions, which are based on the standard fluid dynamical conservation
equations. Since shocks are always optional even when they can occur, we
will compare our results for the relativistic particle acceleration with
those obtained when there is no shock and the flow is globally smooth.

The values of the energy transport parameter $\epsilon$ on the upstream
and downstream sides of the isothermal shock at $r=r_*$ are denoted by
$\epsilon_-$ and $\epsilon_+$, respectively. Note that $\epsilon_- >
\epsilon_+$ as a consequence of the loss of energy through the upper and
lower surfaces of the disk at the shock location. It is important to
emphasize that the drop in $\epsilon$ at the shock has the effect of
altering the transonic structure of the flow in the post-shock region.
Hence, although the post-shock flow must pass through another critical
point and become supersonic before crossing the event horizon, the new
(inner) critical point is {\it not equal} to any of the four roots
computed using the upstream energy transport parameter $\epsilon_-$.
Instead, the new inner critical radius, which we refer to as $\hat
r_{c3}$, must be computed using the {\it downstream} value of the energy
transport parameter, $\epsilon_+$. We point out that the total energy
inflow rate across the horizon, including the rest-mass contribution,
must be positive since no energy can escape from the black hole, and
therefore we require that $c^2 + \epsilon_+ > 0$.

\subsection{Isothermal Shock Jump Conditions}

We shall assume that the escape of the relativistic particles from the
disk results in a negligible amount of mass loss because the Lorentz
factor of the escaping particles is much greater than unity (see
Table~\ref{tbl-2}). This is confirmed ex post facto by comparing the
rate of mass loss, $\dot M_{\rm loss}$, with the accretion rate $\dot
M$. We find that for the models analyzed here, $\dot M_{\rm loss} / \dot
M \lapprox 10^{-3}$, and therefore our assumption of negligible mass
loss is justified. Hence the accretion rate $\dot M$ is conserved as the
gas crosses the shock, which is represented by the condition
\begeq
\Delta \, \dot M
\equiv \ \lim_{\varepsilon \to 0} \dot M(r_*-\varepsilon)
- \dot M(r_*+\varepsilon) \ = \ 0
\ ,
\label{eqnew29}
\fineq
where the symbol ``$\Delta$'' will be used to denote the difference
between post- and pre-shock quantities. The specific angular momentum
$\ell \equiv \dot J/\dot M$ is also conserved throughout the flow, and
therefore we find that
\begin{equation}
\Delta \, \dot J = 0 \ .
\label{eqnew30}
\fineq
Furthermore, the radial momentum transport rate, defined by
\begeq
\dot I \equiv 4 \pi r H (P + \rho \, v^2) \ ,
\label{eqnew31}
\fineq
must remain constant across the shock, and consequently
\begeq
\Delta \, \dot I = 0 \ .
\label{eqnew32}
\fineq
Based on equations~(\ref{eqnew12a}) and (\ref{eqnew30}), we find
that the jump condition for the energy transport rate $\dot E$ is
given by
\begeq
\Delta \, \dot E = \dot M \left({1 \over 2} \, \Delta \, v^2
+ {1 \over \gamma-1} \, \Delta \, a^2\right) \ .
\label{eqnew33}
\fineq
\begfig[t]
\centering
\includegraphics[scale=0.65]{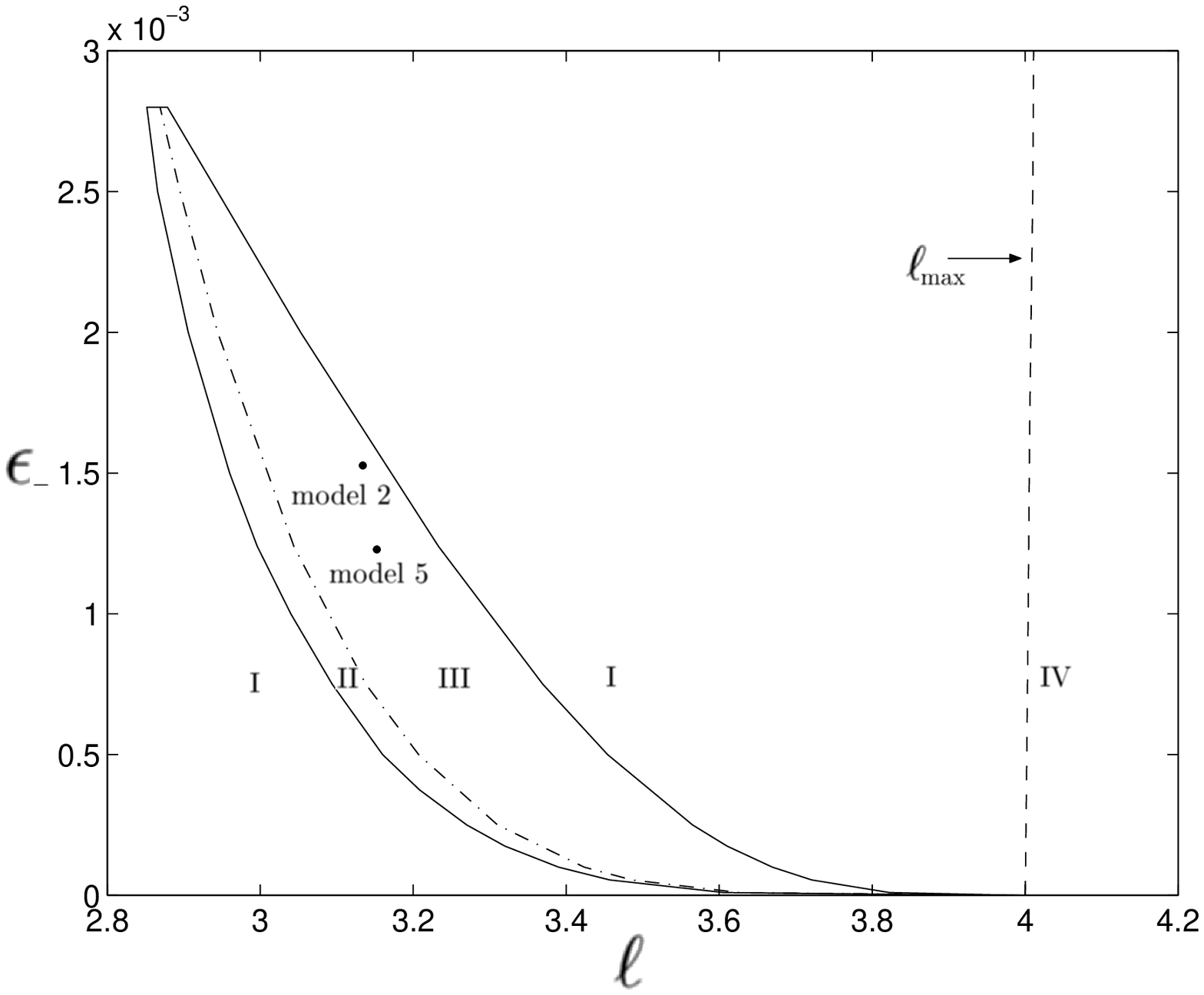}
\caption{Plot of the $(\epsilon_-,\ell)$ parameter space for an ADAF
disk with $\gamma=1.5$. Only smooth flows exist in region~I, and both
shocked and smooth solutions are possible in regions~II and III. When
$\ell > \ell_{\rm max}$ (region~IV), no steady-state dynamical solutions
can be developed. The parameters corresponding to models~2 and 5 are
indicated.}
\label{chpt3_fig5}
\finfig

Equations~(\ref{eqnew4}), (\ref{eqnew8}), and (\ref{eqnew12a}) can
be combined with equations~(\ref{eqnew29}), (\ref{eqnew32}), and
(\ref{eqnew33}) to obtain
\begeqarray
\rho_+ v_+ a_+ & = & \rho_- v_- a_- \ ,
\label{eqnew34} \\
a_+ P_+ - a_- P_- & = & a_- \rho_- v_-^2 - a_+
\rho_+ v_+^2 \ ,
\label{eqnew35} \\
\epsilon_{_+} - \epsilon_{_-} & = &
{v_+^2 - v_-^2 \over 2} + {a_+^2 - a_-^2 \over \gamma-1} \ ,
\label{eqnew36}
\fineqarray
where the subscripts ``-'' and ``+'' refer to quantities measured just
upstream and just downstream from the shock, respectively. In the case
of an isothermal shock, $a_+=a_-$, and therefore the shock jump
conditions reduce to
\begeqarray
\rho_+ v_+ & = & \rho_- v_- \ ,
\label{eqnew37} \\
P_+ - P_- & = & \rho_- v_-^2 - \rho_+ v_+^2 \ ,
\label{eqnew38} \\
\epsilon_+ - \epsilon_- & = & {v_+^2 - v_-^2 \over 2} \ .
\label{eqnew39}
\fineqarray
Combining equations~(\ref{eqnew37}) and (\ref{eqnew38}) and substituting
for the pressure $P$ using equation~(\ref{eqnew9}) yields the velocity
jump condition
\begeq
{v_+ \over v_-} = \gamma^{-1} \, {\cal M}_-^{-2} \ < \ 1 \ ,
\ \ \ \ \
{\cal M}_- \equiv {v_- \over a_-} \ ,
\label{eqnew40}
\fineq
where ${\cal M}_-$ is the incident Mach number of the shock. The
corresponding result for the shock compression ratio $R_*$ is
\begeq
R_* \equiv {\rho_+ \over \rho_-}
= \gamma \, {\cal M}_-^2 \ \ > \ \ 1 \ .
\label{eqnew41}
\fineq
Hence the gas density increases across the shock as expected. Based
on equations~(\ref{eqnew14}) and (\ref{eqnew40}) and the fact that
$a_+ = a_-$ in an isothermal shock, we find that the jump condition
for the entropy parameter $K$ is given by
\begeq
{K_+ \over K_-} = \gamma^{-1} \, {\cal M}_-^{-2} \ < \ 1 \ ,
\label{eqnew42}
\fineq
which indicates that entropy is lost from the disk at the shock location
due to the escape of the particles that form the outflow (jet).

We can also make use of equation~(\ref{eqnew40}) to rewrite the jump
condition for the energy transport parameter (eq.~[\ref{eqnew39}]) as
\begeq
\Delta \epsilon \equiv \epsilon_+ - \epsilon_-
= {v_-^2 \over 2} \, \left({1 \over \gamma^2 {\cal M}_-^4}
-1 \right) \ < \ 0 \ .
\label{eqnew43}
\fineq
The associated rate at which energy escapes from the disk at the
isothermal shock location (the ``shock luminosity'') is given by
(see eq.~[\ref{eqnew12a}])
\begeq
\Lshock \equiv - \Delta \dot E = - \dot M \, \Delta \epsilon
\ \ \propto \ {\rm ergs \ s}^{-1} \ .
\label{eqnew44}
\fineq
Eliminating $\Delta \epsilon$ between equations~(\ref{eqnew43})
and (\ref{eqnew44}) yields the alternative result
\begeq
{\Lshock \over \dot M}
= {v_-^2 \over 2} \, \left(1 - {1 \over \gamma^2 {\cal M}_-^4}
\right) \ .
\label{eqnew45}
\fineq

\subsection{Shock Point Analysis}

For a given value of $\gamma$, it is known that smooth, shock-free
global flow solutions exist only within a restricted region of the
$(\epsilon_-, \ell)$ parameter space, and isothermal shocks can occur
only in a subset of the smooth-flow region. In order for a shock to
exist in the flow, it must be located between two critical points, and
it must also satisfy the jump conditions given by
equations~(\ref{eqnew40}), (\ref{eqnew42}), and (\ref{eqnew43}). The
procedure for determining the disk/shock structure is summarized below.

The process begins with the selection of values for the
fundamental parameters $\epsilon_-$, $\ell$, and $\gamma$. The
values of $\epsilon_-$ and $\ell$ are ultimately constrained by
the observations of a specific object, as discussed in \S~7.
Following Narayan, Kato, \& Honma (1997), we shall assume an
approximate equipartition between the gas and magnetic pressures,
and therefore we set $\gamma=1.5$. The first step in the
determination of the shock location is the computation of the
outer critical point location, $r_{c1}$, using the quartic
equation~(\ref{eqnew26}). The associated values for the critical
velocity, $v_{c1}$, the critical sound speed, $a_{c1}$, and the
critical entropy, $K_{c1}$, are then calculated using
equations~(\ref{eqnew22}), (\ref{eqnew23}), and (\ref{eqnew24}),
respectively. Note that since the flow is adiabatic everywhere in
the pre-shock region, it follows that
\begeq
K_- = K_{c1} \ .
\label{eqnew46}
\fineq
The profiles of the velocity $v(r)$ and the sound speed $a(r)$ in the
pre-shock region can therefore be calculated using a root-finding
procedure based on equation~(\ref{eqnew28}), or, alternatively, by
integrating numerically the wind equation~(\ref{eqnew18}). The next step is
the selection of an initial guess for the shock radius, $r_*$, and the
calculation of the associated shock Mach number ${\cal M}_- \equiv
v_-/a_-$ using the pre-shock dynamical solutions for $v(r)$ and $a(r)$.
Based on the value of ${\cal M}_-$, we can compute the jump in the
entropy parameter $K$ using equation~(\ref{eqnew42}), and consequently we
find that the entropy in the downstream region is given by
\begeq
K_+ = {K_{c1} \over \gamma \, {\cal M}_-^2} \ ,
\label{eqnew47}
\fineq
where we have also used equation~(\ref{eqnew46}).

In order to determine whether the initial guess for $r_*$ is
self-consistent, we employ a second, independent procedure for
calculating the entropy in the downstream region based on the critical
nature of the flow. In this approach, the downstream energy parameter
$\epsilon_+$ is calculated using the jump condition given by
equation~(\ref{eqnew43}), which yields
\begeq
\epsilon_+ = \epsilon_- + {v_-^2 \over 2} \,
\left({1 \over \gamma^2 {\cal M}_-^4}
- 1 \right) \ .
\label{eqnew48}
\fineq
We utilize this value to compute the downstream critical point radius
$\hat r_{c3}$ based on the quartic equation~(\ref{eqnew26}). The
associated values for the critical velocity, $\hat v_{c3}$, the critical
sound speed, $\hat a_{c3}$, and the critical entropy, $\hat K_{c3}$, are
then calculated using equations~(\ref{eqnew22}), (\ref{eqnew23}), and
(\ref{eqnew24}), respectively. The final step is to compare the value of
$\hat K_{c3}$ with that obtained for $K_+$ using
equation~(\ref{eqnew47}). If these two quantities are equal, then the
shock radius $r_*$ is correct and the disk/shock model is therefore
dynamically self-consistent. Otherwise, the value for $r_*$ must be
iterated and the search continued. Roots for $r_*$ can be found only in
certain regions of the $(\epsilon_-,\ell,\gamma)$ parameter space, as
discussed below.

By combining the analysis of the shock location discussed above with the
critical conditions developed in \S~3, we are able to compute the
structure of shocked and shock-free (smooth) disk solutions for a given
set of parameters $\epsilon_-$, $\ell$, and $\gamma$. The resulting
topology of the parameter space is depicted in Figure~\ref{chpt3_fig5}
for the case with $\gamma=1.5$, which is the main focus of this paper.
Within region~I, only smooth flows are possible, and in regions~II and
III both smooth and shocked solutions are available. No global flow
solutions (either smooth or shocked) exist in region~IV, with $\ell >
\ell_{\rm max}$. Inside region~II, one root for the shock radius $r_*$
can be found, and in region~III two shock solutions are available,
although only one actual shock can occur in a given flow. It is unclear
which of the two roots for $r_*$ in region~III is preferred since the
stability properties of the shocks are not completely understood (e.g.,
Chakrabarti 1989a, 1989b; Abramowicz \& Chakrabarti 1990). However, it
is worth noting that the inner shock is always the stronger of the two
because the Mach number diverges as the gas approaches the horizon. The
larger compression ratio associated with the inner location leads to
more efficient particle acceleration and enhanced entropy generation,
and therefore we expect that the inner solution is preferred in nature.
Based on this argument, we will focus on the inner shock location in our
subsequent analysis.

Before we proceed to examine the transport equation for the
relativistic particles, it is important to analyze the asymptotic
solutions obtained for the dynamical variables near the event
horizon and also at large radii. This is a crucial issue since the
nature of the global solutions to the transport equation depends
sensitively on the boundary conditions imposed at large and small
radii. The asymptotic solutions for the dynamical variables near
the event horizon were fully discussed by Becker \& Le (2003). We
shall briefly review their results and then perform a similar
analysis in order to determine the asymptotic properties of the
solutions as $r \to \infty$.

\subsection{Asymptotic Behavior Near the Horizon}

Becker \& Le (2003) demonstrated that the variation of the global
solutions in a viscous ADAF disk becomes purely adiabatic close to
the event horizon, and therefore the asymptotic solutions that
they obtain can be directly applied to our inviscid model. Using
their equations~(47) and (51), we find that the asymptotic
variations of the radial velocity $v$ and the sound speed $a$ near
the horizon are given by
\begeq v^2(r) \propto (r - \rs)^{-1} \ , \ \ \ \ \ \ \ a^2(r)
\propto (r - \rs)^{(1-\gamma)/(1+\gamma)} \ , \ \ \ \ \ r \to \rs
\ . \label{eqnew49} \fineq
The divergence of $v$ as $r \to \rs$ implies that it cannot
represent the standard velocity in the region near the horizon.
However, our dynamical model is consistent with relativity if we
interpret $v$ as the radial component of the four-velocity in this
region (Becker \& Le 2003; Becker \& Subramanian 2005). By
combining these relations with equations~(\ref{eqnew4}),
(\ref{eqnew8}), and (\ref{eqnew10}), we find that the
corresponding results for the asymptotic variations of the disk
half-thickness $H$ and the density $\rho$ become
\begeq H(r) \propto (r - \rs)^{(\gamma+3)/(2\gamma+2)} \ , \ \ \ \
\ \ \ \rho(r) \propto (r - \rs)^{-1/(\gamma+1)} \ , \ \ \ \ \ r
\to \rs \ . \label{eqnew50} \fineq

\subsection{Asymptotic Behavior at Infinity}

We can use the energy transport equation~(\ref{eqnew12a}) and the
entropy equation~(\ref{eqnew14}) to obtain the asymptotic
solutions for $v$ and $a$ at infinity as follows. In the limit $r
\to \infty$, the two dominant terms in equation~(\ref{eqnew12a})
are $\epsilon$ and $a^2/(\gamma-1)$, and therefore we find that
\begeq a^2(r) \to (\gamma-1) \, \epsilon \ , \ \ \ \ \ r \to
\infty \ . \label{eqnew51} \fineq
Recalling that $K$ is constant in the adiabatic upstream flow, we
can combine equations~(\ref{eqnew14}) and (\ref{eqnew51}) to
conclude that the asymptotic variation of the inflow velocity is
given by
\begeq v \propto r^{-5/2} \ , \ \ \ \ \ r \to \infty \ .
\label{eqnew52} \fineq
Finally, based on equations~(\ref{eqnew8}), (\ref{eqnew10}), and
(\ref{eqnew51}), we find that the disk half-height varies as
\begeq H \propto r^{3/2} \ , \ \ \ \ \ r \to \infty \ .
\label{eqnew53} \fineq
We can also combine equations~(\ref{eqnew9}), (\ref{eqnew13}), and
(\ref{eqnew51}) to conclude that the asymptotic behavior of the
density is given by
\begeq \rho \to {\rm const.} \ , \ \ \ \ \ r \to \infty \ .
\label{eqnew54} \fineq

\section{STEADY-STATE PARTICLE ACCELERATION}

Our goal in this paper is to analyze the transport and
acceleration of relativistic particles (ions) in a disk governed
by the dynamical model developed in \S~3 -- 4. For fixed values of
the theory parameters $\epsilon_-$, $\ell$, and $\gamma$, we will
study the transport of particles in disks with and without shocks.
The particle transport model utilized here includes advection,
spatial diffusion, Fermi energization, and particle escape. In
order to maintain consistency with the dynamics of the disk, we
will need to equate the energy escape rate for the relativistic
particles with the ``shock luminosity'' $\Lshock$ given by
equation~(\ref{eqnew44}). Our treatment of Fermi energization
includes both the general compression related to the overall
convergence of the accretion flow, as well as the enhanced
compression that occurs at the shock. In the scenario under
consideration here, the escape of particles from the disk occurs
via vertical spatial diffusion in the tangled magnetic field, as
depicted in Figure~\ref{chpt3_fig2}. To avoid unnecessary
complexity, we will utilize a simplified model in which only the
radial ($r$) component of the spatial particle transport is
treated in detail. In this approach, the diffusion and escape of
the particles in the vertical ($z$) direction is modeled using an
escape-probability formalism. We will treat the relativistic ions
as test particles, meaning that their contribution to the pressure
in the flow is neglected. This assumption is valid provided the
pressure of the relativistic particles turns out to be a small
fraction of the thermal pressure, as discussed in \S~8. The ions
accelerated at the shock are energized via collisions with MHD
waves advected along with the background (thermal) flow, and
therefore the shock width is expected to be comparable to the
magnetic coherence length, $\lambda_{\rm mag}$. This approximation
will be used to determine the rate at which particles escape from
the disk in the vicinity of the shock.

\subsection{Transport Equation}

The Green's function, $\green(E_0,E,r_*,r)$, represents the particle
distribution resulting from the continual injection of $\N0$
particles per second, each with energy $E_0$, from a source located at
the shock radius, $r=r_*$. In a steady-state situation, the Green's
function satisfies the transport equation (Becker 1992)
\begeq
{\partial \green \over \partial t} = 0 = -\vec\nabla \cdot \vec F
- {1 \over 3 E^2} {\partial \over \partial E}
\left(E^3 \vec v \cdot \vec\nabla \green \right)
+ \dot f_{\rm source} - \dot f_{\rm esc} \ ,
\label{eqnew55}
\fineq
where the specific flux $\vec F$ is evaluated using
\begeq
\vec F = -\kappa \vec\nabla\green -
{\vec v E \over 3} {\partial \green \over \partial E} \ ,
\label{eqnew56}
\fineq
and the source and escape terms are given by
\begeq
\dot f_{\rm source} = {\N0 \, \delta(E-E_0) \,
\delta(r-r_*) \over (4 \pi E_0)^2 \, r_* \, H_*} \ , \ \ \ \
\dot f_{\rm esc} = A_0 \, c \, \delta(r-r_*) \, \green \ .
\label{eqnew57}
\fineq
The quantities $E$, $\kappa$, $\vec v$, and $H_* \equiv H(r_*)$
represent the particle energy, the spatial diffusion coefficient, the
vector velocity, and the disk half-thickness at the shock location,
respectively, and the dimensionless parameter $A_0$ determines the rate
of particle escape through the surface of the disk at the shock
location. The vector velocity $\vec v$ has components given by
$\vec v = v_r \hat r + v_z \hat z + v_\phi \, \hat\phi$, where
$v=-v_r$ is the positive inflow speed.

The total number and energy densities of the relativistic particles,
denoted by $n_r$ and $U_r$, respectively, are related to the Green's
function via
\begeq
n_r(r) = \int_0^\infty 4 \pi E^2 \, \green \, dE \ , \ \ \ \ \
U_r(r) = \int_0^\infty 4 \pi E^3 \, \green \, dE \ ,
\label{eqnew58}
\fineq
which determine the normalization of $\green$. Equations~(\ref{eqnew55})
and (\ref{eqnew56}) can be combined to obtain the alternative form
\begeq
\vec v \cdot \vec\nabla \green = {\vec\nabla \cdot
\vec v \over 3} \, E \, {\partial \green \over \partial E} +
\vec\nabla \cdot \left(\kappa \, \vec\nabla \green
\right) + \dot f_{\rm source} - \dot f_{\rm esc} \ ,
\label{eqnew59}
\fineq
where the left-hand side represents the co-moving (advective) time
derivative and the terms on the right-hand side describe first-order
Fermi acceleration, spatial diffusion, the particle source, and the
escape of particles from the disk at the shock location, respectively.
Note that escape and particle injection are localized to the shock
radius due to the presence of the $\delta$-functions in
equations~(\ref{eqnew57}). Our focus here is on the first-order Fermi
acceleration of relativistic particles at a standing shock in an
accretion disk, and therefore equation~(\ref{eqnew59}) does not include
second-order Fermi processes that may also occur in the flow due to MHD
turbulence (e.g., Schlickeiser 1989a,b; Subramanian, Becker, \& Kazanas
1999).

Under the assumption of cylindrical symmetry, equations~(\ref{eqnew57})
and (\ref{eqnew59}) can be rewritten as
\begeqarray
v_r {\partial \green \over \partial r}
+ v_z {\partial \green \over \partial z}
&-& {1 \over 3}\left[{1 \over r} {\partial \over \partial r}
\left(r v_r\right) + {d v_z \over dz}\right]
E {\partial \green \over \partial E}
- {1 \over r} {\partial \over \partial r}
\left(r \kappa {\partial \green \over \partial r}\right)
\nonumber \\
&=& {\N0 \, \delta(E-E_0) \, \delta(r-r_*) \over
(4 \pi E_0)^2 \, r_* \, H_*} - A_0 \, c \, \delta(r-r_*) \, \green \ ,
\label{eqnew60}
\fineqarray
where the escape of particles from the disk is described by the final
term on the right-hand side. In Appendix~A, we demonstrate that the
vertically integrated transport equation is given by (see
eq.~[\ref{Aeq9}])
\begeqarray
H v_r {\partial \green \over \partial r}
&=& {1 \over 3 r} {\partial \over \partial r} \left(
r H v_r \right) E {\partial \green \over \partial E}
+ {1 \over r} {\partial \over \partial r}
\left(r H \kappa {\partial \green \over \partial r}\right)
\nonumber \\
&+& {\N0 \, \delta(E-E_0) \, \delta(r-r_*) \over
(4 \pi E_0)^2 \, r_*} - A_0 \, c \, H_* \, \green \,
\delta(r-r_*) \ ,
\label{eqnew61}
\fineqarray
where the symbols $\green$, $v_r$, and $\kappa$ represent vertically
averaged quantities. We establish in Appendix~B that within the context
of our one-dimensional spatial model, the dimensionless escape parameter
$A_0$ appearing in equation~(\ref{eqnew61}) is given by (see
eqs.~[\ref{Beq8}] and [\ref{Beq10}])
\begeq
A_0 = \left({3 \, \kappa_* \over c \, H_*}\right)^2 \ < \ 1 \ ,
\label{eqnew62}
\fineq
where $\kappa_* \equiv (\kappa_- + \kappa_+)/2$ denotes the mean value
of the diffusion coefficient at the shock location. The condition $A_0 <
1$ is required for the validity of the diffusive picture we have
employed in our model for the vertical transport.

\subsection{Number and Energy Densities}

The energy moments of the Green's function, $I_n(r)$, are defined by
\begeq
I_n(r) \equiv \int_0^\infty 4 \pi E^n \, \green \, dE \ ,
\label{eqnew63}
\fineq
so that (cf. eqs.~[\ref{eqnew58}])
\begeq
n_r(r) = I_2(r) \ , \ \ \ \ \
U_r(r) = I_3(r) \ .
\label{eqnew64}
\fineq
By operating on equation~(\ref{eqnew61}) with $\int_0^\infty 4
\pi E^n \, dE$ and integrating by parts once, we find that the
function $I_n$ satisfies the differential equation
\begeqarray
H v_r {d I_n \over d r}
&=& - \left({n+1 \over 3}\right) {I_n \over r} \, {d \over d r}
\left(r H v_r \right)
+ {1 \over r} {d \over d r}
\left(r H \kappa {d I_n \over d r}\right)
\nonumber \\
&+& {\N0 \, E_0^{n-2} \, \delta(r-r_*) \over
4 \pi \, r_*} - A_0 \, c \, H_* \, I_n \,
\delta(r-r_*) \ ,
\label{eqnew65}
\fineqarray
which can be expressed in the flux-conservation form
\begeq
{d \over d r} \left(4 \pi r H F_n \right)
= 4 \pi r H \left[\left({2-n \over 3}\right) v {d I_n \over d r}
+ {\N0 \, E_0^{n-2} \,  \delta(r-r_*) \over
4 \pi r_* H_*} - A_0 \, c \, \delta(r-r_*) \, I_n\right] \ ,
\label{eqnew66}
\fineq
where $4 \pi r H F_n$ represents the rate of transport of the
$n$th moment, and the flux $F_n$ is defined by
\begeq
F_n \equiv - \left({n+1 \over 3}\right) v \, I_n - \kappa \,
{d I_n \over d r} \ ,
\label{eqnew67}
\fineq
and $v = - v_r$ denotes the positive inflow speed.

In order to close the system of equations and solve for the relativistic
particle number and energy densities $I_2(r)$ and $I_3(r)$ using
equation~(\ref{eqnew65}), we must also specify the radial variation of
the diffusion coefficient $\kappa$. The behavior of $\kappa$ can be
constrained by considering the fundamental physical principles governing
accretion onto a black hole. First, we note that near the event horizon,
particles are swept into the black hole at the speed of light, and
therefore advection must dominate over diffusion. This condition applies
to both the thermal (background) and the nonthermal (relativistic)
particles. Second, we note that in the outer region ($r \to \infty$),
diffusion is expected to dominate over advection. Focusing on the flux
equation for the particle number density $n_r$, obtained by setting
$n=2$ in equation~(\ref{eqnew67}), we can employ scale analysis to
conclude based on our physical constraints that
\begeq
\lim_{r \to \rs}{\kappa(r) \over (r - \rs) \, v(r)} = 0 \ ,
\ \ \ \ \ \ \ \ \ \ \
\lim_{r \to \infty}{r \, v(r) \over \kappa(r)} = 0 \ .
\label{eqnew68}
\fineq
The precise functional form for the spatial variation of $\kappa$ is not
completely understood in the accretion disk environment. In order to
obtain a mathematically tractable set of equations with a reasonable
physical behavior, we shall utilize the general form
\begeq
\kappa(r) = \kappa_0 \, v(r) \, \rs \left({r \over \rs} -1
\right)^\alpha \ ,
\label{eqnew69}
\fineq
where $\kappa_0$ and $\alpha$ are dimensionless constants. Due to the
appearance of the inflow speed $v$ in equation~(\ref{eqnew69}), we note
that $\kappa$ exhibits a jump at the shock. This is expected on physical
grounds since the MHD waves that scatter the ions are swept along with
the thermal background flow, and therefore they should also experience
a density compression at the shock.

As discussed above, close to the event horizon, inward advection at the
speed of light must dominate over outward diffusion. Conversely, in the
outer region, we expect that diffusion will dominate over advection as
$r \to \infty$. By combining equation~(\ref{eqnew69}) with the asymptotic
velocity variations expressed by equations~(\ref{eqnew49}) and
(\ref{eqnew52}), we find that the conditions given by
equations~(\ref{eqnew68}) are satisfied if $\alpha > 1$, and in our work
we set $\alpha=2$. Note that the escape parameter $A_0$ is related to
$\kappa_0$ via equation~(\ref{eqnew62}), which can be combined with
equation~(\ref{eqnew69}) to write
\begeq
A_0 = \left({3 \kappa_0 \, v_* \rs \over c H_*}\right)^2
\left({r_* \over \rs} - 1\right)^4 < \, 1 \ ,
\label{eqnew70}
\fineq
where $v_* \equiv (v_- + v_+)/2$ represents the mean velocity at
the shock location $r=r_*$. The value of the diffusion parameter
$\kappa_0$ is constrained by the inequality in
equation~(\ref{eqnew70}). In \S~7 we demonstrate that $\kappa_0$
can be computed for a given source based on energy conservation
considerations. With the introduction of equations~(\ref{eqnew69})
and (\ref{eqnew70}), we have completely defined all of the
quantities in the transport equation, and we can now solve for the
number and energy densities of the relativistic particles. The
particle distribution Green's function $\green$ and its
applications will be discussed in a separate paper.

\section{SOLUTIONS FOR THE ENERGY MOMENTS}

Once the disk/shock dynamics have been computed based on the
selected values for the free parameters $\epsilon_-$, $\ell,$ and
$\gamma$ using the results in \S\S~3 and 4, the associated
solutions for the number and energy densities of the relativistic
particles in the disk can be obtained by solving
equation~(\ref{eqnew65}). In the case of the number density, $n_r
= I_2$, an exact solution can be derived based on the linear
first-order differential equation describing the conservation of
particle flux. However, in order to understand the variation of
the energy density, $U_r = I_3$, we must numerically integrate a
second-order equation. The solutions obtained below are applied in
\S~7 to model the outflows observed in M87 and \SgrA.

\subsection{Relativistic Particle Number Density}

The equation governing the transport of the particle number density,
$n_r$, is obtained by setting $n=2$ in equation~(\ref{eqnew66}), which
yields
\begeq
{d \dot N_r \over d r}
= \N0 \, \delta(r-r_*)
- 4 \pi r_* H_* A_0 \, c \, n_r \, \delta(r-r_*) \ ,
\label{eqnew71}
\fineq
where the relativistic particle transport rate, $\dot N_r(r)$, is
defined by (cf. eq.~[\ref{eqnew67}])
\begeq
\dot N_r(r) \equiv - 4 \pi r H \left(v \, n_r + \kappa \,
{d n_r \over d r}\right) \ \propto \ {\rm s}^{-1} \ ,
\label{eqnew72}
\fineq
and $\dot N_r > 0$ if the transport is in the outward direction. Since
the source is located at the shock, there are two spatial domains of
interest in our calculation of the particle transport, namely domain~I
($r > r_*$), and domain~II ($r < r_*$). Note that the number density
$n_r(r)$ must be continuous at $r=r_*$ in order to avoid generating an
infinite diffusive flux according to equation~(\ref{eqnew72}). Away from
the shock location, $r \neq r_*$, and therefore equation~(\ref{eqnew71})
reduces to $\dot N_r = \rm const.$, which reflects the fact that
particle injection and escape are localized at the shock. We can
therefore write
\begeq
\dot N_r(r) = \cases{
\NI \ , & $r > r_*$ \ , \cr
\NII \ , & $r < r_*$ \ , \cr
}
\label{eqnew73}
\fineq
where the constant $\NI > 0$ denotes the rate at which particles
are transported outward, radially, from the source location, and the
constant $\NII < 0$ represents the rate at which particles are
transported inward towards the event horizon.

The magnitude of the jump in the particle transport rate at the shock is
obtained by integrating equation~(\ref{eqnew71}) with respect to radius
in a very small region around $r=r_*$, which yields
\begin{equation}
\NI - \NII = \N0 - \Nesc \ ,
\label{eqnew74}
\fineq
where
\begin{equation}
\Nesc \equiv 4 \pi r_* H_* A_0 \, c \, n_*
% \ \propto \ {\rm s}^{-1}
\label{eqnew75}
\fineq
represents the (positive) rate at which particles escape from the disk
at the shock location to form the outflow (jet), and $n_* \equiv
n_r(r_*)$. If no shock is present in the flow, then $A_0=0$ and
therefore $\Nesc$=0. Note that the discontinuity in $\dot N_r$ at the
shock produces a jump in the derivative $dn_r/dr$ via
equation~(\ref{eqnew72}).

We can rewrite equation~(\ref{eqnew72}) for the number density in
the form
\begeq
{dn_r \over dr} + {v \over \kappa} \, n_r = - {\dot N_r \over 4
\pi r H \kappa} \ ,
\label{eqnew76}
\fineq
which is a linear, first-order differential equation for $n_r(r)$.
Using the standard integrating factor technique and employing
equation~(\ref{eqnew69}) for $\kappa$ yields the exact solution
\begeq
n_r(r) = e^{- J(r)}
\left[n_* - {\dot N_r(r) \over 4 \pi} \int_{r_*}^r
{e^{J(r')} \over
r' H \kappa} \, dr' \right] \ ,
\label{eqnew77}
\fineq
where $\dot N_r(r)$ is given by equation~(\ref{eqnew73}) and the
function $J(r)$ is defined by
\begeq
J(r) \equiv \int_{r_*}^r v \kappa^{-1} \, dr'
= \kappa_0^{-1} \left[\left({r_* \over \rs} - 1\right)^{-1}
- \left({r \over \rs} - 1\right)^{-1}\right] \ .
\label{eqnew78}
\fineq
According to equation~(\ref{eqnew77}), $n_r(r)$ is continuous at the
shock/source location as required. Far from the black hole, diffusion
dominates the particle transport, and therefore $n_r$ should vanish as $
r \to \infty$. In order to ensure this behavior, we must have
\begeq
n_* = \NI \, \CI \ ,
\label{eqnew79}
\fineq
where
\begeq
\CI \equiv {1 \over 4 \pi} \int_{r_*}^\infty
{e^{J(r')} \over
r' H \kappa} \, dr' \ .
\label{eqnew80}
\fineq
Furthermore, in order to avoid exponential divergence of $n_r$ as $r \to
\rs$ in domain~II, we also require that
\begeq
n_* = - \NII \, \CII \ ,
\label{eqnew81}
\fineq
where
\begeq
\CII \equiv {1 \over 4 \pi} \int_{\rs}^{r_*}
{e^{J(r')} \over
r' H \kappa} \, dr' \ .
\label{eqnew82}
\fineq

By combining equations~(\ref{eqnew74}), (\ref{eqnew75}),
(\ref{eqnew79}), and (\ref{eqnew81}), we can develop explicit
expressions for the quantities $n_*$, $\NI$, $\NII$, and $\Nesc$ based
on the values of $r_*$ and $\N0$ and the profiles of the inflow velocity
$v(r)$ and the diffusion coefficient $\kappa(r)$. The results obtained
are
\begeqarray
n_* & = & {\N0 \over \CI^{-1} + \CII^{-1} + 4 \pi
r_* H_* A_0 \, c} \ ,
\nonumber \\
\NI & = & {n_* \over \CI} \ ,
\label{eqnew83} \\
\NII & = & - {n_* \over \CII} \ ,
\nonumber \\
\Nesc & = & 4 \pi r_* H_* A_0 \, c \, n_* \ .
\nonumber
\fineqarray
These relations, along with equation~(\ref{eqnew77}), complete the
formal solution for the relativistic particle number density $n_r(r)$.
The solution is valid in both shocked and shock-free disks (the
shock-free case is treated by setting $A_0=0$). When a shock is present,
the particle escape rate $\Nesc$ is proportional to $\N0$ but
is independent of $E_0$ by virtue of equations~(\ref{eqnew83}).

It is interesting to examine the asymptotic variation of $n_r$ near the
event horizon and also at large distances from the black hole. Far from
the hole, advection is negligible and the particle transport in the disk
is dominated by outward-bound diffusion. In this case we can use
equation~(\ref{eqnew76}) to conclude that
\begeq
{dn_r \over dr} \to - {\NI \over 4 \pi r H \kappa} \ ,
\ \ \ \ \ \ \ r \to \infty \ ,
\label{eqnew84}
\fineq
where we have used the fact that $\dot N_r = \NI$ for $r > r_*$. By
combining equations~(\ref{eqnew69}) and (\ref{eqnew84}) with the
asymptotic relations given by equations~(\ref{eqnew52}) and
(\ref{eqnew53}), we find upon integration that
\begeq
n_r(r) \propto {1 \over r} \ ,
\ \ \ \ \ \ \ r \to \infty \ .
\label{eqnew85}
\fineq
In order to study the behavior of $n_r$ near the event horizon, we take
the limit as $r \to \rs$ in equation~(\ref{eqnew77}), obtaining after
some algebra
\begeq
n_r(r) \to - {\NII \over 4 \pi r H v} \ , \ \ \ \ \ r \to \rs \ ,
\label{eqnew86}
\fineq
where we have set $\dot N_r = \NII$. Comparing this relation with
equation~(\ref{eqnew4}), we find that
\begeq
n_r(r) \ \propto \ \rho(r)
\ , \ \ \ \ \ \ \ \quad r \to \rs \ ,
\label{eqnew87}
\fineq
where $\rho$ is the density of the background (thermal) gas.
Equation~(\ref{eqnew87}) is a natural consequence of the fact
that the particle transport near the horizon is dominated by
inward-bound advection. We can also combine equations~(\ref{eqnew50})
and (\ref{eqnew87}) to obtain the explicit asymptotic form
\begeq
n_r(r) \ \propto (r - \rs)^{-1/(\gamma+1)}
\ , \ \ \ \ \ \ \ \quad r \to \rs \ .
\label{eqnew88}
\fineq

\subsection{Relativistic Particle Energy Density}

The differential equation satisfied by the relativistic particle energy
density, $U_r = I_3$, is obtained by setting $n=3$ in
equation~(\ref{eqnew65}), which yields
\begeqarray
H v_r {d U_r \over d r}
&=& - {4 \over 3} \, {U_r \over r} \, {d \over d r}
\left(r H v_r \right)
+ {1 \over r} {d \over d r}
\left(r H \kappa {d U_r \over d r}\right)
\nonumber \\
&+& {\N0 \, E_0 \, \delta(r-r_*) \over
4 \pi \, r_*} - A_0 \, c \, H_*
\, U_r \, \delta(r-r_*) \ .
\label{eqnew89}
\fineqarray
By analogy with equations~(\ref{eqnew71}) and (\ref{eqnew72}), we can
recast this expression in the flux-conservation form
\begeq
{d \dot E_r \over d r}
= 4 \pi r H \left[- {v \over 3} {d U_r \over dr}
+ {\N0 \, E_0 \, \delta(r-r_*) \over 4 \pi r_* H_*}
- A_0 \, c \, U_r \, \delta(r-r_*) \right] \ ,
\label{eqnew90}
\fineq
where the relativistic particle energy transport rate, $\dot E_r(r)$,
is defined by
\begeq
\dot E_r(r) \equiv - 4 \pi r H \left({4 \over 3} \, v \, U_r
+ \kappa \, {d U_r \over d r}\right) \ \propto \ {\rm ergs \ s}^{-1}
\ ,
\label{eqnew91}
\fineq
and $\dot E_r > 0$ for outwardly directed transport. Note that unlike
the number transport rate $\dot N_r$, the energy transport rate $\dot
E_r$ does {\it not} remain constant within domains~I and II due to the
appearance of the first term on the right-hand side of
equation~(\ref{eqnew90}), which expresses the compressional work done on
the relativistic particles by the background flow.

Although the energy density $U_r$ must be continuous at the shock/source
location in order to avoid generating an infinite diffusive flux, the
derivative $dU_r/dr$ displays a discontinuity at $r=r_*$, which is
related to the jump in the energy transport rate via
equation~(\ref{eqnew91}). By integrating equation~(\ref{eqnew90}) in a
very small region around $r=r_*$, we find that
\begeq
\Delta \dot E_r = \Lesc - \N0 \, E_0 \ ,
\label{eqnew92}
\fineq
where
\begeq
\Lesc \equiv 4 \pi r_* H_* A_0 \, c \, U_*
\ \propto \ {\rm ergs \ s}^{-1}
\label{eqnew93}
\fineq
denotes the rate of escape of energy from the disk into the outflow
(jet) at the shock location, and $U_* \equiv U_r(r_*)$. If no shock is
present, then $A_0=0$ and therefore $\Lesc=0$. We remind the reader that
the symbol ``$\Delta$'' refers to the difference between post- and
pre-shock quantities (see eq.~[\ref{eqnew29}]).
Equations~(\ref{eqnew91}) and (\ref{eqnew92}) can be combined to show
that the derivative jump is given by
\begeq
\Delta \left(\kappa \, {dU_r \over dr}\right)
= {\N0 \, E_0 - \Lesc \over 4 \pi r_* H_*}
- {4 \over 3} \, U_* \, \Delta v \ .
\label{eqnew94}
\fineq

The differential equation~(\ref{eqnew89}) governing the relativistic
particle energy density is second-order in radius, and therefore we will
need to establish two boundary conditions in order to solve for
$U_r(r)$. These can be obtained by analyzing the behavior of $U_r$ close
to the event horizon and at large distances from the black hole. Far
from the hole, advection is negligible and the particle transport in the
disk is dominated by outward-bound diffusion. In this regime, Fermi
acceleration is negligible, and consequently we find that $U_r \propto
n_r$. We can therefore use equation~(\ref{eqnew85}) to conclude that
\begeq
U_r(r) \propto {1 \over r} \ ,
\ \ \ \ \ \ \ r \to \infty \ .
\label{eqnew95}
\fineq
Close to the event horizon, the particle transport is dominated by
advection, and therefore $U_r$ and $n_r$ obey the standard adiabatic
relation
\begeq
U_r \propto n_r^{4/3} \ , \ \ \ \ \ \ \ r \to \rs \ .
\label{eqnew96}
\fineq
Combining this result with equations~(\ref{eqnew87}) then yields
\begeq
U_r \propto (r - \rs)^{-4/(3\gamma+3)} \ ,
\ \ \ \ \ \ \ r \to \rs \ .
\label{eqnew97}
\fineq

The global solution for $U_r(r)$ can now be expressed as
\begeq
U_r(r) = \cases{
A \ \QI(r) \ , & $r > r_*$ \ , \cr
%
%\phantom{stuff} \cr
%
B \ \QII(r) \ , & $r < r_*$ \ , \cr
}
\label{eqnew98}
\fineq
where $A$ and $B$ are constants and the functions $\QI(r)$ and $\QII(r)$
satisfy the homogeneous differential equation (see eq.~[\ref{eqnew89}])
\begeq
H v_r {d Q \over d r}
= - {4 \over 3} \, {Q \over r} \, {d \over d r}
\left(r H v_r \right)
+ {1 \over r} {d \over d r}
\left(r H \kappa {d Q \over d r}\right) \ .
\label{eqnew99}
\fineq
along with the boundary conditions (see eqs.~[\ref{eqnew95}] and
[\ref{eqnew97}])
\begeq
\QI(r_{\rm out}) = \left({r_{\rm out} \over \rs}\right)^{-1}
\ , \ \ \ \ \ \ \
\QII(r_{\rm in}) = \left({r_{\rm in} \over
\rs}-1\right)^{-4/(3\gamma + 3)} \ ,
\label{eqnew100}
\fineq
where $r_{\rm in}$ and $r_{\rm out}$ denote the radii at which the inner
and outer boundary conditions are applied, respectively. The constants
$A$ and $B$ are computed by requiring that $U_r$ be continuous at
$r=r_*$ and that the derivative $dU_r/dr$ satisfy the jump condition
given by equation~(\ref{eqnew94}). The results obtained are
\begeqarray
A &=& B \, {\QII \over \QI} \Bigg|_{r=r_*} \ ,
\label{eqnew101} \\
B &=& {\N0 \, E_0 \over 4 \pi r_* H_*}
\Bigg[{4 \over 3} \, (v_- - v_+) \, \QII + \QII' \kappa_+
- {\QII \QI' \kappa_- \over \QI} + A_0 \, c \, \QII
\Bigg]^{-1} \Bigg|_{r=r_*} \ ,
\label{eqnew102}
\fineqarray
where primes denote differentiation with respect to radius. The
solutions for the functions $Q_{\rm I}(r)$ and $Q_{\rm II}(r)$ are
obtained by integrating numerically equation~(\ref{eqnew99})
subject to the boundary conditions given by
equations~(\ref{eqnew100}). Once the constants $A$ and $B$ are
computed using equations~(\ref{eqnew101}) and (\ref{eqnew102}),
the global solution for $U_r(r)$ is evaluated using
equation~(\ref{eqnew98}). The solution applies whether or not a
shock is present in the flow. The shock-free case is treated by
setting $A_0=0$. This completes the solution procedure for the
relativistic particle energy density. The results derived in this
section are used in \S~7 to model the outflows observed in M87 and
\SgrA.

\section{ASTROPHYSICAL APPLICATIONS}

Our goal is to determine the properties of the integrated
disk/shock/outflow model based on the observed values for the black hole
mass $M$, the mass accretion rate $\dot M$, and the jet kinetic power
$\Ljet$ associated with a given source. The fundamental free parameters
for the theoretical model are $\epsilon_-$, $\ell$, and $\gamma$. Since
we set $\gamma = 1.5$ in order to represent an approximate equipartition
between the gas and magnetic pressures (e.g., Narayan, Kato, \& Honma
1997), only $\epsilon_-$ and $\ell$ remain to be determined. Here we
describe how global energy conservation considerations can be used to
solve for the various theoretical parameters in the model based on
observations.

\subsection{Energy Conservation Conditions}

Once the values of $M$, $\dot M$, and $\Ljet$ have been specified for
a source based on observations, we select a value for the free parameter
$\epsilon_-$ and then compute $\ell$ by satisfying the relation
\begeq
\Lshock = \Ljet \ ,
\label{eqnew103}
\fineq
where $\Lshock$ is the shock luminosity given by
equation~(\ref{eqnew45}). This result ensures that the jump in the
energy transport rate at the isothermal shock location is equal to
the observed jet kinetic luminosity. The procedure for determining
$\ell$ also includes solving for the shock location and the
critical structure using results from \S\S~3 and 4. The velocity
profile $v(r)$ is computed either by numerically integrating
equation~(\ref{eqnew18}) or by using a root-finding procedure
based on equation~(\ref{eqnew28}), and the associated solution for
the adiabatic sound speed $a(r)$ is obtained using
equation~(\ref{eqnew12a}).

After the velocity profile has been determined, we can compute the
number and energy density distributions for the relativistic particles
in the disk using equations~(\ref{eqnew77}) and (\ref{eqnew98}),
respectively. This requires the specification of the injection energy of
the seed particles $E_0$ as well as their injection rate $\N0$. We set
the injection energy using $E_0 = 0.002\,$ergs, which corresponds to an
injected Lorentz factor $\Gamma_0 \equiv E_0 / (m_p \, c^2) \sim 1.3$,
where $m_p$ is the proton mass. Particles injected with energy $E_0$ are
subsequently accelerated to much higher energies due to repeated shock
crossings. We find that the speed of the injected particles, $v_0 = c \,
(1-\Gamma_0^{-2})^{1/2}$, is about three to four times higher than the
mean ion thermal velocity at the shock location, $v_{\rm rms} = (3 k
T_*/m_p)^{1/2}$, where $T_*$ is the ion temperature at the shock. The
seed particles are therefore picked up from the high-energy tail of the
Maxwellian distribution for the thermal ions. With $E_0$ specified, we can
compute the particle injection rate $\N0$ using the energy conservation
condition
\begeq
\dot N_0 \, E_0 = \Lshock \ ,
\label{eqnew104}
\fineq
which ensures that the rate at which energy is injected into the flow in
the form of the relativistic seed particles is equal to the energy loss
rate for the background gas at the isothermal shock location.

In order to maintain agreement between the transport model and the
observations, we must also require that the rate at which particle
energy {\it escapes} from the disk due to vertical diffusion is equal to
the observed jet power. This condition can be written as
\begeq
\Lesc = \Ljet \ ,
\label{eqnew105}
\fineq
where $\Lesc$ is the energy escape rate given by
equation~(\ref{eqnew93}). The escape constant, $A_0$, appearing in the
transport equation is independent of the particle energy in our model,
and consequently the escaping particles will have exactly the same mean
energy as those in the disk at the shock location. The mean energy of
the escaping particles is therefore given by
\begeq
\Eesc \equiv {U_* \over n_*} \ ,
\label{eqnew106}
\fineq
where $n_*$ and $U_*$ denote the number and energy densities of the
relativistic particles at the shock location, respectively. Hence
$\Eesc$ is proportional to $E_0$ but it is independent of $\N0$.
We note that equations~(\ref{eqnew75}), (\ref{eqnew93}), and
(\ref{eqnew106}) can be
combined to show that
\begeq
\Lesc = \Nesc \, \Eesc \ ,
\label{eqnew107}
\fineq
where $\Nesc$ is the particle escape rate (eq.~[\ref{eqnew75}]). By
satisfying equations~(\ref{eqnew103}), (\ref{eqnew104}), and
(\ref{eqnew105}), we ensure that energy is properly conserved in our
model. Taken together, these relations allow us to solve for the various
theoretical parameters based on observational values for $M$, $\dot M$,
and $\Ljet$, as explained below.

\begin{table}[t]
\centering
\scriptsize
\caption{Disk structure parameters
\label{tbl-1}}
% Tabular environment goes AFTER the caption!
\begin{tabular}{cccccccccccccc}
\hline \hline
Model
& $\epsilon_{_-}$
& $\ell$
& $\epsilon_{_+}$
& $r_{c1}$
& $r_{c3}$
& $r_*$
& $\hat r_{c3}$
& $H_*$
& ${\cal M}_*$
& $R_*$
& $T_*$     \\ \hline
1          %
&0.001600  %
&3.1040    %
&-0.005676 %
&93.177    %
&5.478     %
&19.624    %
&5.798     %
&10.369    %
&1.094     %
&1.795     %
&1.28 \\%
2          %
&0.001527  %
&3.1340    %
&-0.005746 %
&98.524    %
&5.379     %
&21.654    %
&5.659     %
&11.544
&1.125     %
&1.897     %
&1.16 \\%
3          %
&0.001400  %
&3.1765    %
&-0.005875 %
&109.781   %
&5.252     %
&24.500    %
&5.487     %
&13.154
&1.170     %
&2.052     %
&1.01 \\ %
4          %
&0.001240  %
&3.1280    %
&-0.008723 %
&131.204   %
&5.408     %
&14.073    %
&5.850     %
&6.819      %
&1.113     %
&1.857     %
&1.65 \\%
5          %
&0.001229  %
&3.1524    %
&-0.008749 %
&131.874   %
&5.329     %
&15.583    %
&5.723     %
&7.672     %
&1.146     %
&1.970     %
&1.49 \\ %
6          %
&0.001200  %
&3.1756    %
&-0.008778 %
&135.192   %
&5.260     %
&16.950    %
&5.614     %
&8.434     %
&1.177     %
&2.076     %
&1.36   \\ %
\hline \\
\end{tabular} \\
\raggedright
All quantities are expressed in gravitational units ($GM=c=1$)
except $T_*$ which is in units of $10^{11}\,$K.

\end{table}

\subsection{Model Parameters}

Our simulations of the disk structure and particle transport in M87 and
\SgrA are based on various published observational estimates for $M$,
$\dot M$, and $\Ljet$. For M87, we set $M = 3 \times 10^9 {\rm \
M_\odot}$ (e.g., Ford et al. 1994), $\dot M = 1.3 \times 10^{-1} {\rm \
M_\odot \ yr^{-1}}$ (e.g., Reynolds et al. 1996), and $\Ljet = 5.5
\times 10^{43}\,{\rm ergs \ s^{-1}}$ (e.g., Reynolds et al. 1996;
Bicknell \& Begelman 1996; Owen, Eilek, \& Kassim 2000). For \SgrA, we
use the values $M = 2.6 \times 10^6 {\rm \ M_\odot}$ (e.g., Sch\"odel et
al. 2002) and $\dot M = 8.8 \times 10^{-7} {\rm \ M_\odot \ yr^{-1}}$
(e.g., Yuan, Markoff, \& Falcke 2002; Quataert 2003). Although the
kinetic luminosity of the jet in \SgrA is rather uncertain (see, e.g.,
Yuan 2000; Yuan et al. 2002), we will adopt the value quoted by Falcke
\& Biermann (1999), and therefore we set $\Ljet = 5 \times 10^{38} {\rm
ergs \ s^{-1}}$.

We study both shocked and shock-free solutions spanning the
computational domain between the inner radius $r_{\rm in} = 2.001$ and
the outer radius $r_{\rm out}=5000$, where $r_{\rm in}$ and $r_{\rm
out}$ are the radii at which the boundary conditions are applied (see
eqs.~[\ref{eqnew100}]). Six different accretion/shock scenarios are
explored in detail, with the values for the various parameters
$\epsilon_-$, $\ell$, $\epsilon_+$, $r_{c1}$, $r_{c3}$, $r_*$, $\hat
r_{c3}$, $H_*$, ${\cal M}_*$, $R_*$, and $T_*$ reported in
Table~\ref{tbl-1}. Models~1, 2, and 3 are associated with M87, while
models 4, 5, and 6 are used to study \SgrA. In our numerical examples,
we utilize natural gravitational units ($GM=c=1$ and $\rs=2$), except as
noted. Based on the observational values for $\dot M$ and $\Ljet$
associated with the two sources, we can use equations~(\ref{eqnew44})
and (\ref{eqnew103}) to conclude that $\Delta \epsilon = -0.007$ for M87
and $\Delta \epsilon = -0.01$ for \SgrA. These results are consistent
with the values for $\epsilon_-$ and $\epsilon_+$ reported in
Table~\ref{tbl-1}, and therefore $\Lshock = \Ljet$ as required (see
eq.~[\ref{eqnew103}]).

Next we use the energy conservation condition $\Lesc = \Ljet$
(eq.~[\ref{eqnew105}]) to determine the value of the diffusion
constant $\kappa_0$ (eq.~[\ref{eqnew69}]) for a shocked disk. In
Figure~\ref{chpt3_fig6}, we plot $\Lesc/\Ljet$, $\Gamma_{\rm esc}$, and
$\Nesc/\N0$ as functions of $\kappa_0$ for the M87 and \SgrA parameters,
where $\Gamma_{\rm esc} \equiv \Eesc / (m_p \, c^2)$ is the mean
Lorentz factor of the escaping particles. The treatment of energy
conservation in our disk/shock model is self-consistent when the
condition $\Lesc / \Ljet = 1$ is satisfied, which corresponds to
specific values of $\kappa_0$ as indicated in
Figures~\ref{chpt3_fig6}{\it a} and \ref{chpt3_fig6}{\it d}. We find
that two $\kappa_0$ roots exist for models~3 and 6, one root is possible
for models~2 and 5, and no roots exist for models~1 and 4. Hence the
values of $\epsilon_-$ associated with models~2 and 5 represent the {\it
maximum possible values} for $\epsilon_-$ that yield self-consistent
solutions based on the M87 and \SgrA data, respectively. For
illustrative purposes, we shall focus on the details of the disk
structure and particle transport obtained in models 2 and 5.

\begfig[t]
\centering
\includegraphics[scale=0.78]{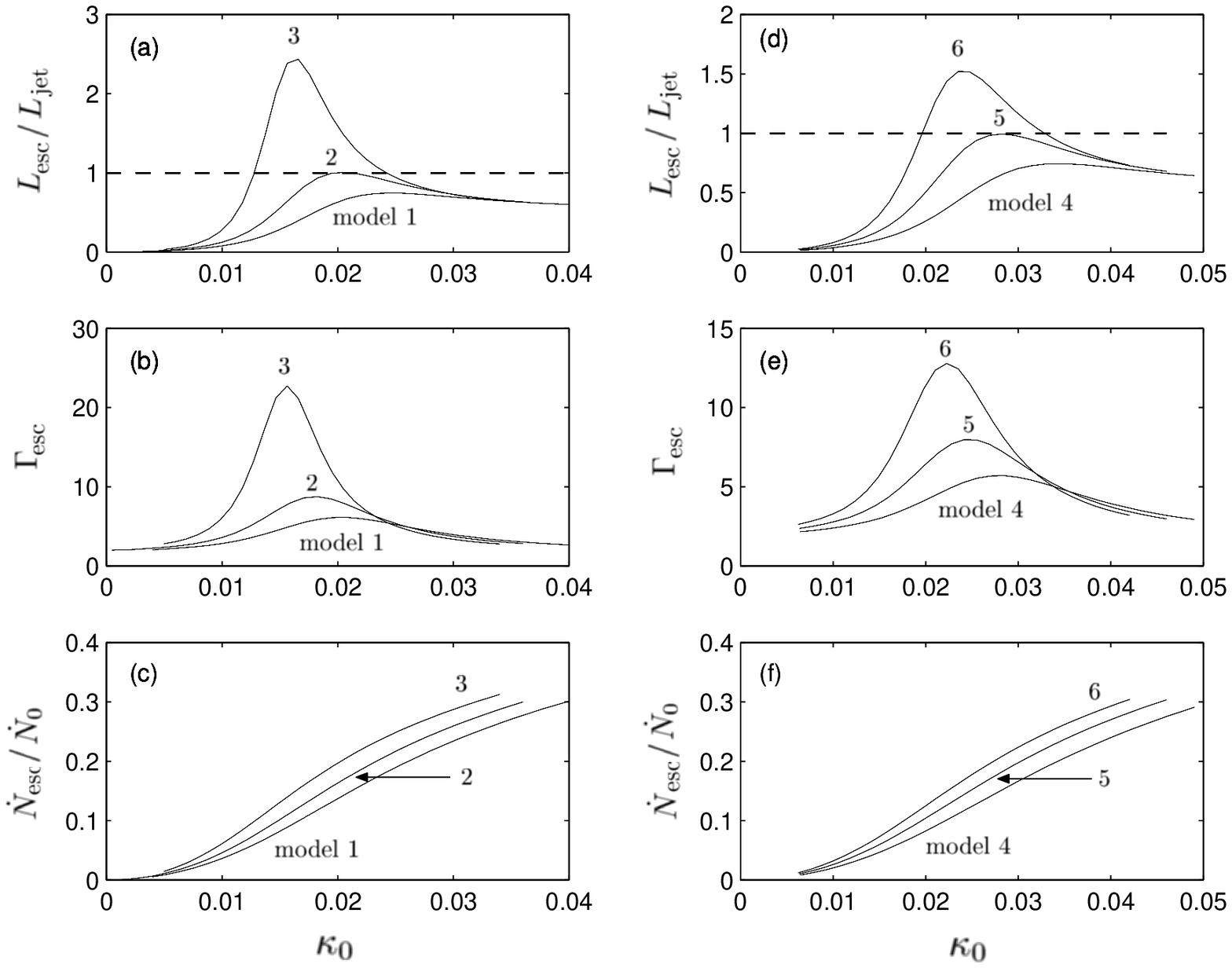}
\scriptsize
\caption{Quantities $\Lesc/\Ljet$, $\Gamma_{\rm esc}$, and $\Nesc/\N0$
for a shocked disk plotted as functions of $\kappa_0$. Panels ({\it a}),
({\it b}), and ({\it c}) correspond to the M87 parameters (models~1, 2
and 3), and panels ({\it d}), ({\it e}), and ({\it f}) correspond to the
\SgrA parameters (models~4, 5 and 6). The model numbers are indicated
for each curve. See the discussion in the text.} \label{chpt3_fig6}
\finfig
\begin{table}[t]
\centering
\scriptsize
\caption{Transport equation parameters
\label{tbl-2}}
% Tabular environment goes AFTER the caption!
\begin{tabular}{cccccccccccccc}
\hline \hline
Model
& $\N0$
& $\underline{\NI}$
& $\kappa_0$
& $\kappa_*$
& $A_0$
& $n_*$
& $U_*$
& $\underline{\Nesc}$
& $\underline{\Eesc}$
& $\Gamma_{\rm esc}$
\\

& (s$^{-1}$)
& $\NII$
&
&
&
& (cm$^{-3}$)
& (ergs cm$^{-3}$)
& $\N0$
& $E_0$
& \\ \hline
2
& 2.75 $\times 10^{46}$
& -0.18
& 0.02044
& 0.427877
& 0.0124
& $2.01 \times 10^4$
& $2.39 \times 10^2$
& $0.17$
& 5.95
& 7.92
\\
5
& 2.51 $\times 10^{41}$
& -0.15
& 0.02819
& 0.321414
& 0.0158
& $4.33 \times 10^5$
& $4.71 \times 10^3$
& $0.18$
& 5.45
& 7.26
\\
\hline \\
\end{tabular}

\raggedright
All quantities are expressed in gravitational units ($GM=c=1$)
except as noted.

\end{table}

\subsection{Disk Structure and Particle Transport}

In order to illustrate the importance of the shock for the acceleration
of high-energy particles, we shall examine the structure of the
accretion disk either with and without a shock based on the values of
the upstream parameters $\epsilon_-$ and $\ell$ utilized in models~2 and
5 (see Table~\ref{tbl-1}). In Figures~\ref{chpt3_fig7}{\it a} and
\ref{chpt3_fig7}{\it b} we plot the inflow speed $v(r)$ and the
adiabatic sound speed $a(r)$ for the shocked and smooth (shock-free)
solutions associated with models~2 and 5, respectively. Since we are
working within the isothermal shock scenario, the sound speed $a$ is
continuous at the shock location. In Figures~\ref{chpt3_fig7}{\it c} and
\ref{chpt3_fig7}{\it d} we plot the specific internal energy
$U/\rho = (\gamma-1)^{-1}k T/m_p$ for the background (thermal)
gas along with the specific gravitational potential (binding) energy
$GM/(r-\rs)$ as functions of radius for $\gamma=1.5$. These results
demonstrate that the gas is marginally bound in the absence of a shock,
and strongly bound when a shock is present. The increased binding of the
thermal gas in the disk results from the escape of energy in the
outflow, which reduces the sound speed compared with the shock-free
case. The enhanced cooling allows the accretion to proceed, thereby
removing one of the major objections to the original ADAF scenario
(Narayan \& Yi 1994, 1995). We emphasize that these new results
represent the first fully self-consistent calculations of the structure
of an ADAF disk coupled with a shock-driven outflow, hence extending the
heuristic work of Blandford \& Begelman (1999) and Becker, Subramanian,
\& Kazanas (2001).

\begfig[t]
\centering
\includegraphics[scale=.75]{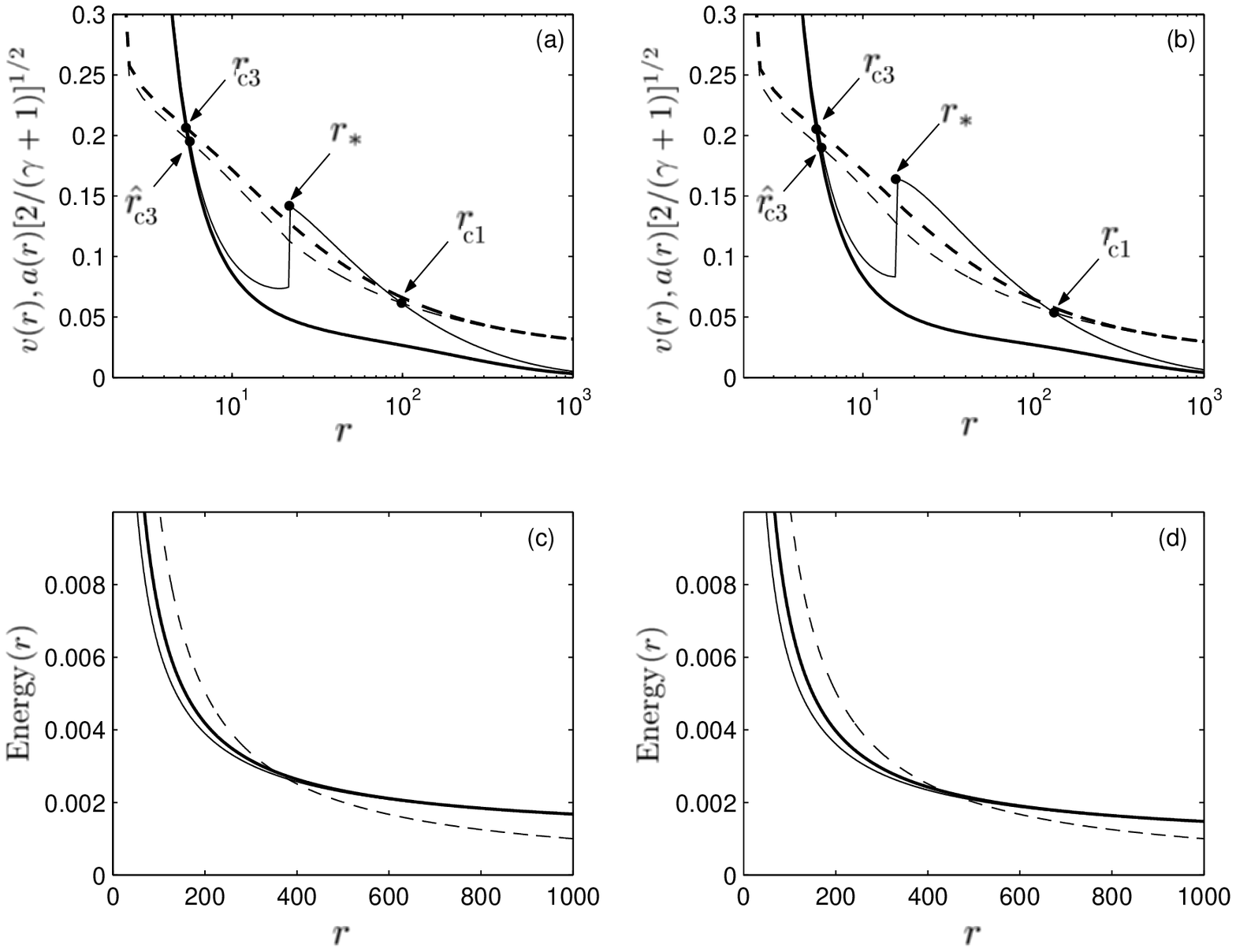}
\scriptsize
\caption{The velocity $v(r)$ ({\it solid lines}) and sound speed $a(r)
\, [2/(\gamma+1)]^{1/2}$ ({\it dashed lines}) are plotted in units of
$c$ for ({\it a}) model~2 and ({\it b}) model~5. The curves cross at the
critical points. Also plotted are the specific internal energy $U/\rho$
({\it solid lines}) and the specific gravitational potential energy
$GM/(r-\rs)$ ({\it dashed lines}), both in units of $c^2$, for ({\it c})
model~2 and ({\it d}) model~5. The shocked and shock-free solutions are
denoted by the thin and heavy lines, respectively.}
\label{chpt3_fig7}
\finfig

Next we study the solutions obtained for the relativistic particle
number and energy density distributions in the disk based on the flow
structures associated with models~2 and 5. The related transport
parameters are listed in Table~\ref{tbl-2}. In Figures~\ref{chpt3_fig10}
and \ref{chpt3_fig11} we plot the global number and energy density
distributions obtained in a shocked disk using the model~2 and 5
parameters, respectively. We also include the corresponding results
obtained in a shock-free (smooth) disk for the same values of the
upstream energy transport rate $\epsilon_-$ and the specific angular
momentum $\ell$. In each case the densities decrease monotonically with
increasing radius. The increase near the horizon is a consequence of
advection, while the decline as $r \to \infty$ reflects the fact that
the particles injected at the shock have a very small chance of
diffusing to large distances from the black hole. Note
that the shocked disk has a lower value for the number density $n_r$ at
all radii as a consequence of particle escape. However, the shocked disk
also displays a {\it higher} value for the energy density $U_r$, which
reflects the central role of shock in accelerating the relativistic test
particles.

The kinks in the energy and number density distributions at the shock
radius $r=r_*$ indicated in Figures~\ref{chpt3_fig10} and
\ref{chpt3_fig11} reflect the derivative jump conditions given by
equations~(\ref{eqnew74}) and (\ref{eqnew94}). The values for the ratios
$\NI/\NII$ and $\Nesc/\dot N_0$ reported in Table~\ref{tbl-2} indicate
that most of the injected particles are advected into the black hole,
with $\sim 20$\% escaping to form the outflow (see
Figs.~\ref{chpt3_fig6}{\it c} and \ref{chpt3_fig6}{\it f}). In order to
validate the accuracy of the numerical solutions for $n_r(r)$ and
$U_r(r)$, we also compare the profiles obtained with the asymptotic
relations developed in \S~6. We demonstrate in Figure~\ref{chpt3_fig8}
(model 2) and Figure~\ref{chpt3_fig9} (model 5) that the solutions for
both $n_r(r)$ and $U_r(r)$ agree closely with the asymptotic expressions
given by equations~(\ref{eqnew88}) and (\ref{eqnew97}) for small radii
and by equations~(\ref{eqnew85}) and (\ref{eqnew95}) for large radii.
Note that the values reported by Le \& Becker (2004) for $n_*$ and $U_*$
were expressed in incorrect units and are given correctly in our
Table~\ref{tbl-2}.

\begfig[t]
\centering
\includegraphics[scale=0.7]{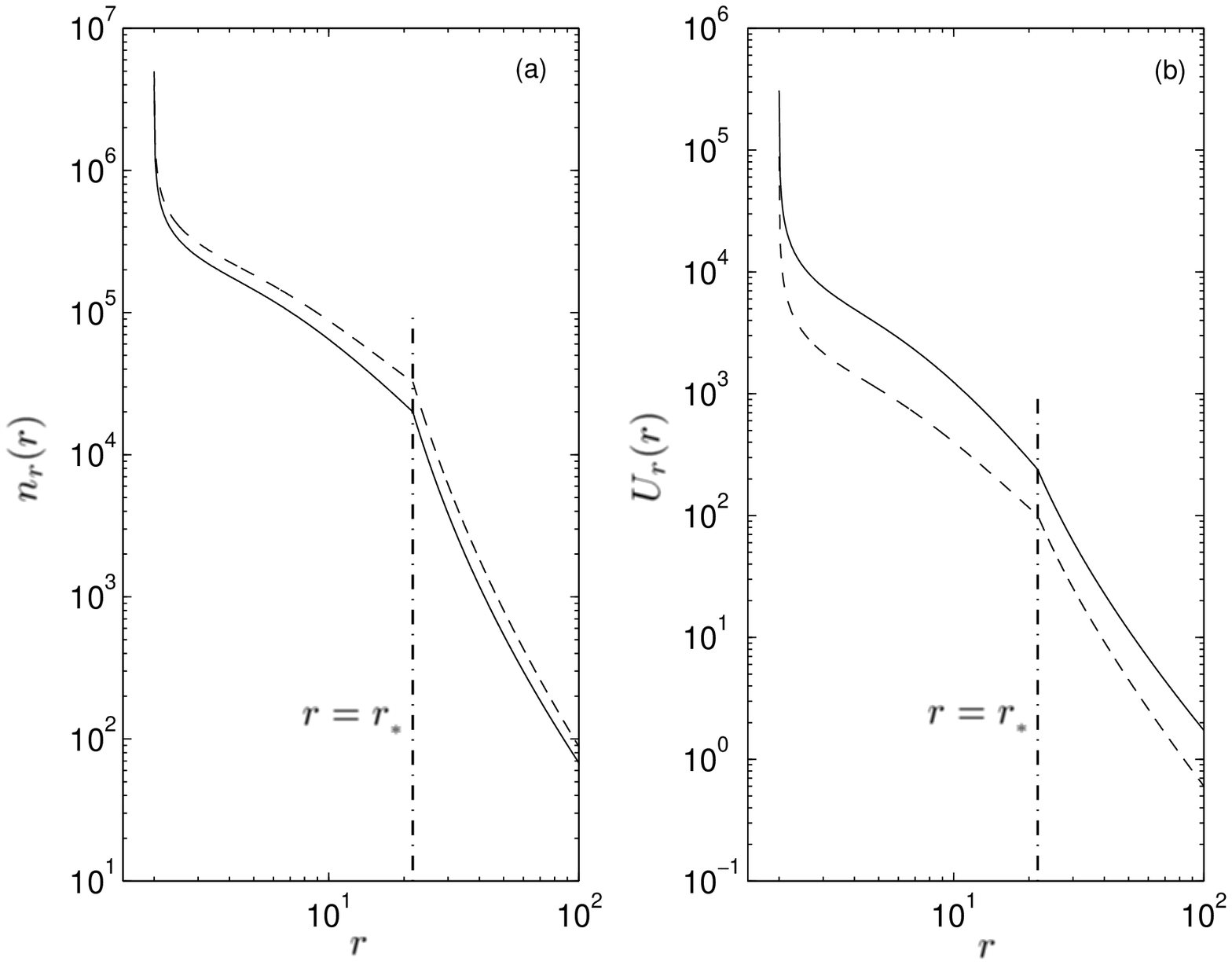}
\scriptsize
\caption{Global solutions for the relativistic number density ({\it a})
and the relativistic energy density ({\it b}) computed using the model~2
parameters. The solid and dashed curves correspond to disks with and
without shocks, respectively. Note that the number density is higher in
the smooth (shock-free) disk due to the absence of particle escape.
Conversely, the energy density is higher in the shocked disk due to the
enhanced particle acceleration occurring at the shock.}
\label{chpt3_fig10}
\finfig
\begfig[t]
\centering
\includegraphics[scale=0.7]{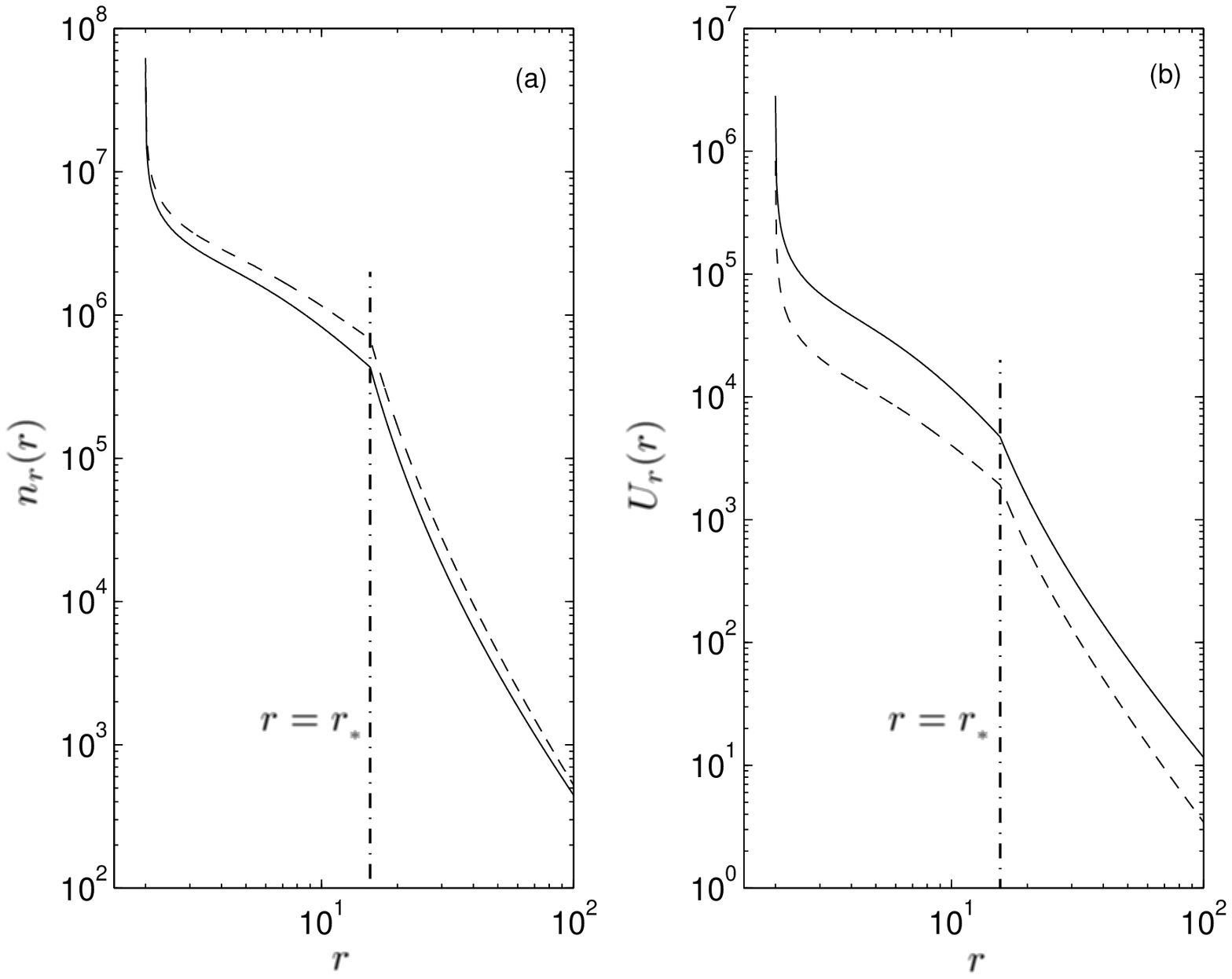}
\scriptsize
\caption{Same as Figure~\ref{chpt3_fig10} but for model~5.}
\label{chpt3_fig11}
\finfig
\begfig[t]
\centering
\includegraphics[scale=0.7]{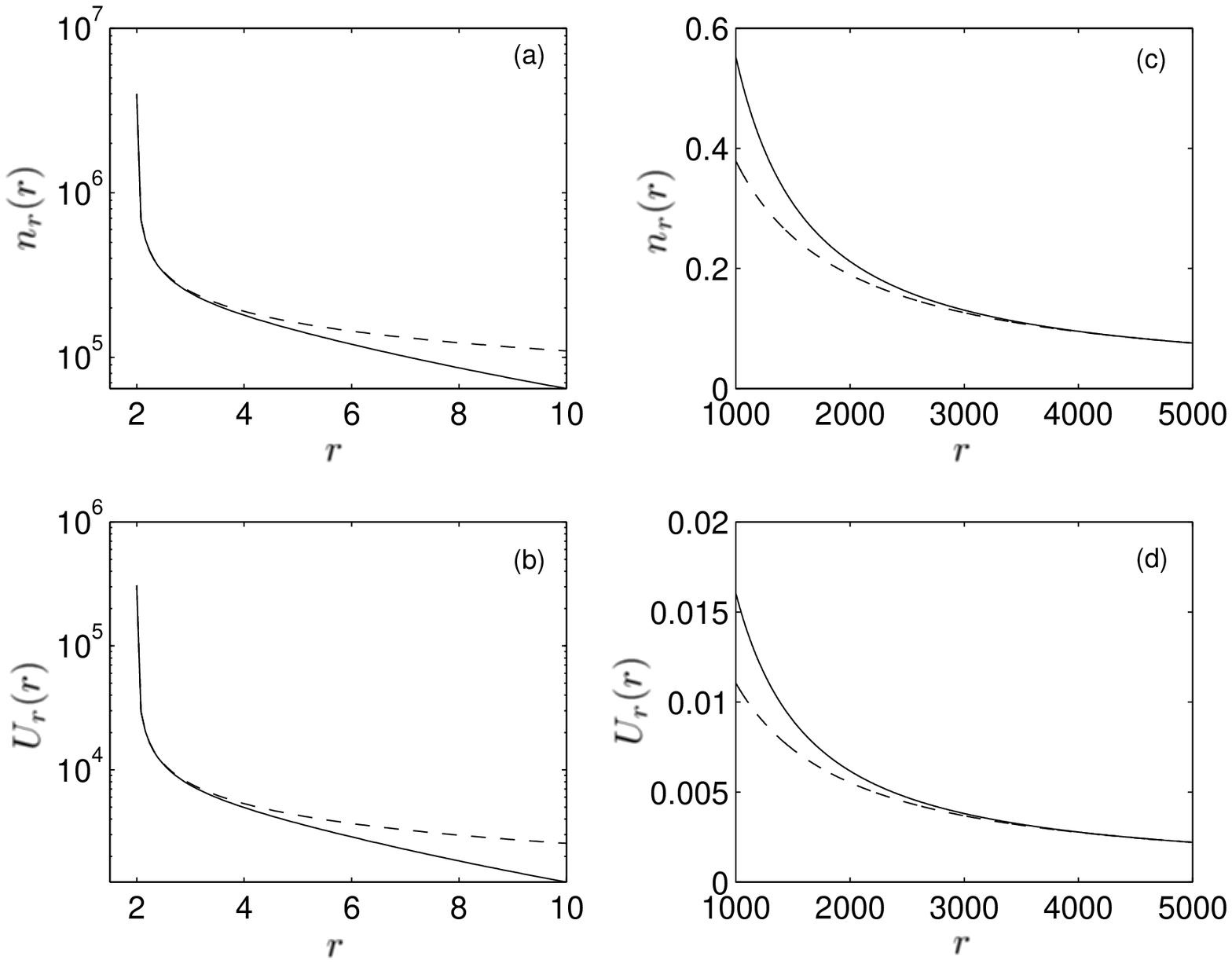}
\scriptsize
\caption{Plots of the numerical solutions for $n_r(r)$ and $U_r(r)$
({\it solid lines}) computed using the model~2 parameters in a shocked
disk are compared with the asymptotic expressions ({\it dashed lines})
close to the event horizon ({\it a} and {\it b}) and at large radii
({\it c} and {\it d}). See the discussion in the text.}

\label{chpt3_fig8}
\finfig
\begfig[t]
\centering
\includegraphics[scale=0.7]{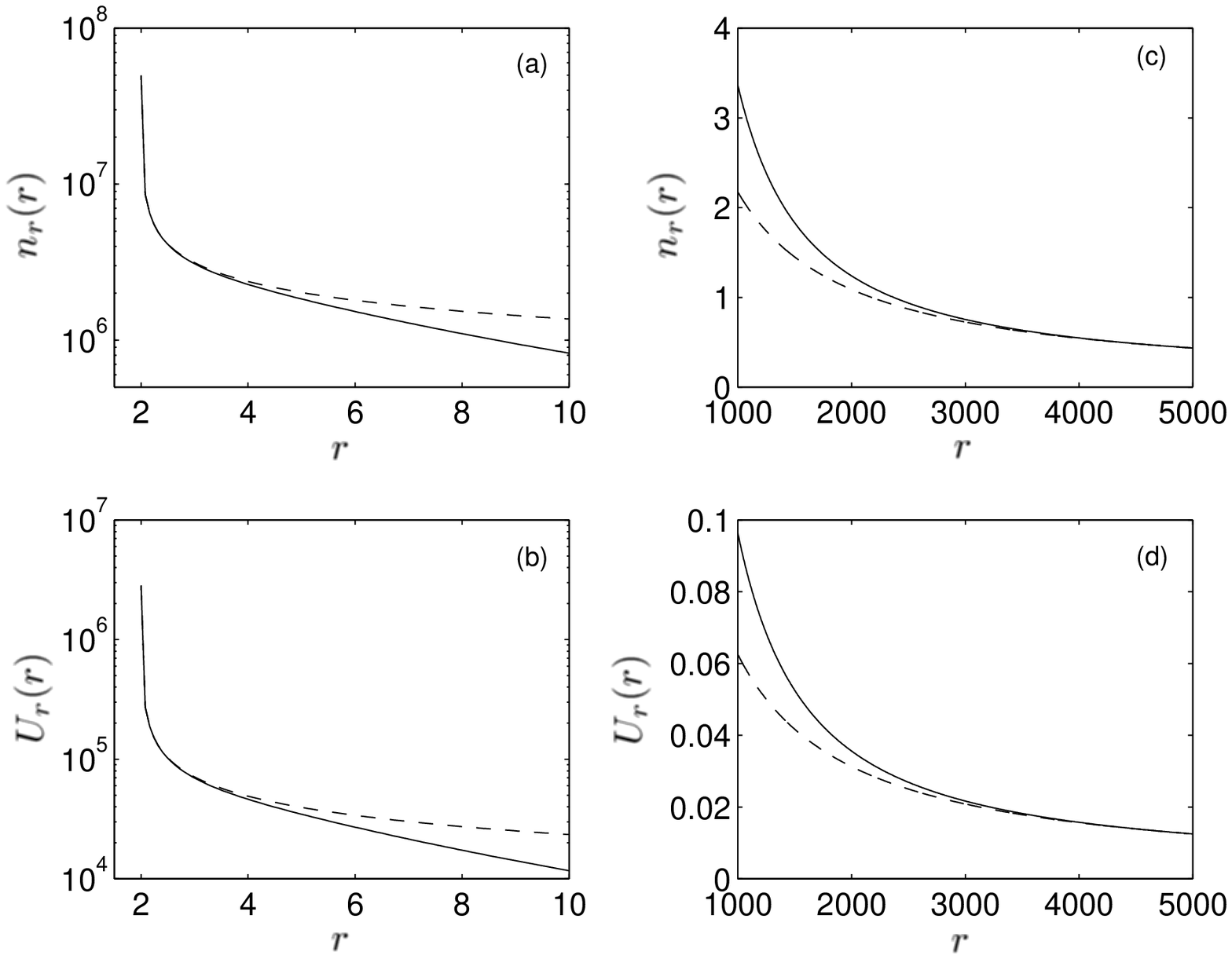}
\scriptsize
\caption{Same as Figure~\ref{chpt3_fig8} but for model~5.}

\label{chpt3_fig9}
\finfig

\subsection{Jet Formation in M87 and \SgrA}

The mean energy of the relativistic particles in the disk is given by
(cf. eq.~[\ref{eqnew106}])
\begeq
\langle E \rangle \equiv {U_r(r) \over n_r(r)}
\ ,
\label{eqnew108}
\fineq
so that $\langle E \rangle = \Eesc$ at $r=r_*$. In
Figure~\ref{chpt3_fig12} we plot the mean energy as a function of radius
in shocked and shock-free disks based on the parameters used for
models~2 and 5. The results demonstrate that when a shock is present in
the flow, the relativistic particle energy is boosted by a factor of
$\sim 5-6$ at the shock location. By contrast, we find that in the
shock-free models with the same values for $\epsilon_-$, $\ell$, and
$\kappa_0$, the energy is boosted by a factor of only $\sim 1.4-1.5$.
This clearly demonstrates the essential role of the shock in efficiently
accelerating particles up to very high energies, far above the energy
required to escape from the disk. Note that close to the event horizon,
the mean energy of the relativistic particles is further enhanced by the
strong compression of the accretion flow, as indicated by the sharp
increase in $\langle E \rangle$ as $r \to \rs$.

The material in the outflow is initially ejected from the disk in the
vicinity of the shock as a hot plasma which cools as it expands, with
its outward acceleration powered by the pressure gradient in the
surrounding plasma. Based on our results for models~2 and 5, we find
that the shock/jet locations are given by $r_* \sim 22$ and $r_* \sim
16$ for M87 and \SgrA, respectively. The terminal (asymptotic) Lorentz
factor of the jet, $\Gamma_\infty$, can be estimated by writing
\begeq
\Gamma_\infty = \Gamma_{\rm esc} = {\Eesc \over m_p c^2} \ ,
\label{eqnew109}
\fineq
which is based on the assumption that the jet starts off ``slow'' and
``hot'' and subsequently expands to become ``fast'' and ``cold.''
Adopting the $\Gamma_{\rm esc}$ values listed in Table~\ref{tbl-2} for M87 and
\SgrA, we obtain $\Gamma_\infty=7.92$ (see Fig.~\ref{chpt3_fig6}{\it b})
and $\Gamma_\infty=7.26$ (see Fig.~\ref{chpt3_fig6}{\it e}),
respectively.

\begfig[t]
\centering
\includegraphics[scale=0.7]{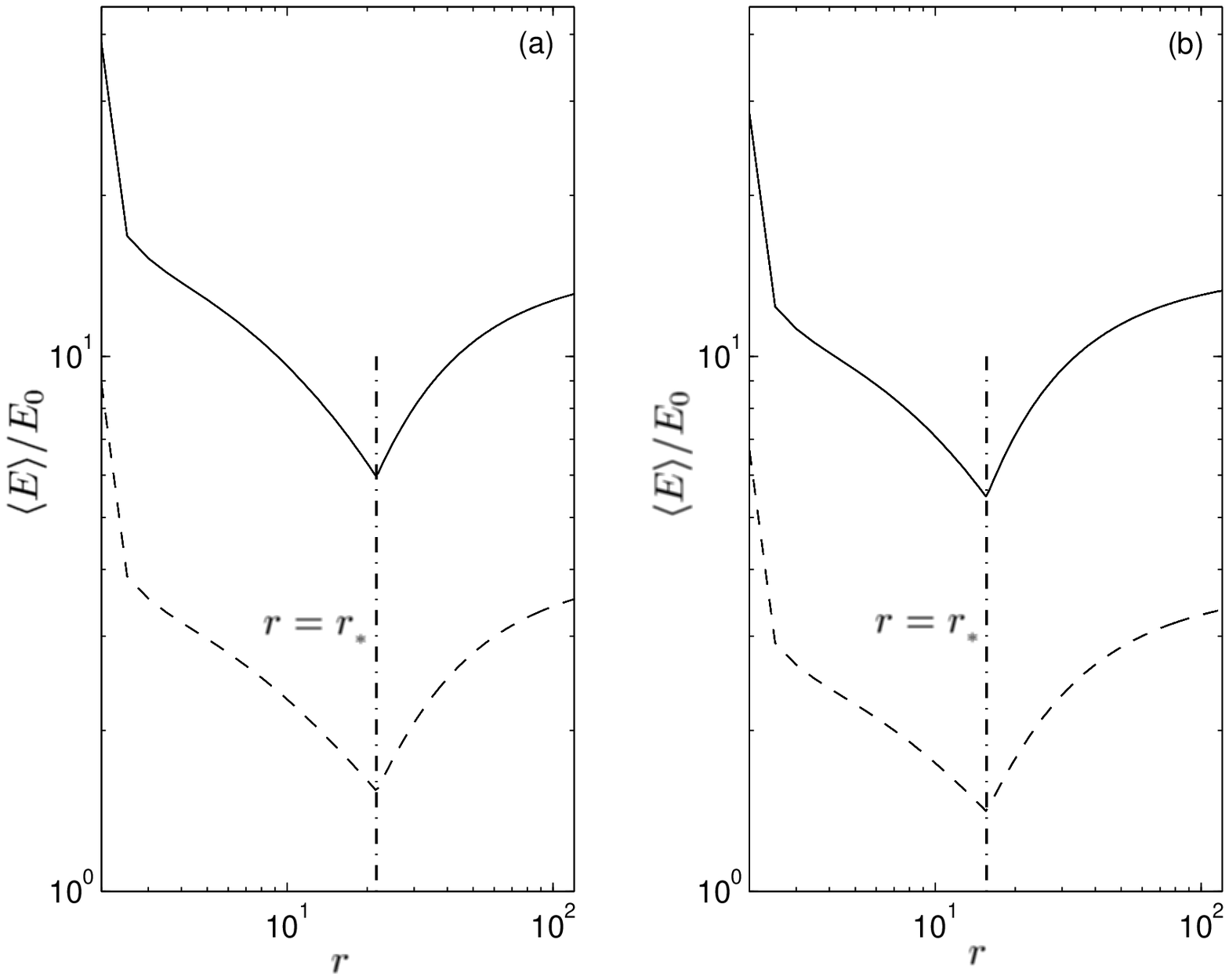}
\scriptsize
\caption{Mean energy of the relativistic particles in the disk,
$\langle E \rangle\equiv U_r(r)/n_r(r)$ (eq.~[\ref{eqnew108}]), plotted
in units of the injection energy $E_0$ as a function of radius for ({\it
a}) model~2 and ({\it b}) model~5. Results are indicated for both
shocked ({\it solid lines}) and shock-free ({\it dashed lines})
disk structures.}
\label{chpt3_fig12}
\finfig

We can now compare our model predictions for the shock/jet location and
the asymptotic Lorentz factor with the observations of M87 and \SgrA.
According to Biretta, Junor, \& Livio (2002), the M87 jet forms in a
region no larger than $\sim 30$ gravitational radii from the black hole,
which agrees rather well with our predicted shock/jet location $r_* \sim
22$ for this source. Turning now to the asymptotic (terminal) Lorentz
factor, we note that Biretta, Sparks, \& Macchetto (1999) estimated
$\Gamma_{\infty} \geq 6$ for the M87 jet, which is comparable to the
result $\Gamma_{\rm \infty} = 7.92$ obtained using our model. In the
case of \SgrA, our model indicates that the shock forms at $r_* \sim 16$
which is fairly close to the value suggested by Yuan (2000). However,
future observational work will be needed to test our prediction for the
asymptotic Lorentz factor of \SgrA, since no reliable observational
estimate for that quantity is currently available.

\subsection{Radiative Losses from the Jet}

It is still unclear whether the outflows observed to emanate from
many radio-loud systems containing black holes are composed of an
electron-proton plasma or electron-positron pairs, or a mixture of
both. Whichever is the case, the particles must maintain
sufficient energy during their journey from the nucleus in order
to power the observed radio emission, unless some form of
reacceleration takes place along the way, due to shocks
propagating along the jet (e.g., Atoyan \& Dermer 2004a).
Proton-electron outflows, such as those studied here, have a
distinct advantage in this regard since most of the kinetic power
is carried by the ions, which do not radiate much and are not
strongly coupled to the electrons under the typical conditions in
a jet (e.g., Felten 1968; Felten, Arp, \& Lynds 1970; Anyakoha et
al. 1987; Aharonian 2002). We therefore suggest that if the
observed outflows are proton-driven, then they may be powered
directly by the shock acceleration mechanism operating in the
disk, with no requirement for additional in situ reacceleration in
the jet. In this section we confirm this conjecture by considering
the energy losses experienced by the protons in the outflow. The
ions in the jet lose energy via two distinct channels, namely (1)
direct radiative losses due to the production of synchrotron and
inverse-Compton emission, and (2) indirect radiative losses via
Coulomb coupling with the electrons. We will evaluate these two
possibilities by estimating the corresponding cooling timescales
for the outflows in M87 and \SgrA and comparing the results with
the jet propagation timescales for these sources.

The energy loss rate due to the production of synchrotron and
inverse-Compton emission by the relativistic protons escaping from the
disk with mean energy $\Gamma_{\rm esc} m_p c^2$ is given by (see
eqs.~[7.17] and [7.18] from Rybicki \& Lightman 1979)
\begeq
\left(dE \over dt\right)\Bigg|_{\rm rad}
= {4 \, \sig c \, \Gamma^2_{\rm esc} \over 3}
\left(m_e \over m_p\right)^2
\left(U_B + U_{\rm ph} \right)
\ ,
\label{eqnew110}
\fineq
where $U_{\rm ph}$ and $U_B = B^2/(8 \pi)$ denote the energy densities
of the soft radiation and the magnetic field with strength $B$, respectively.
The associated energy loss timescale is therefore
\begeq
t_{\rm rad} \equiv {\Gamma_{\rm esc} m_p c^2 \over
~(dE / dt)\big|_{\rm rad}}
= {3 \, m_p \, c \over 4 \, \sig \, \Gamma_{\rm esc}}
\left(m_p \over m_e\right)^2
\left(U_B + U_{\rm ph} \right)^{-1}
\ .
\label{eqnew111}
\fineq
In our application to M87, we take $B \sim 0.1\,$G based on estimates
from Biretta, Stern, \& Harris (1991), and we set $\Gamma_{\rm esc} \sim
8$ (see Table~2, model 2). Assuming equipartition between the magnetic
field and the soft radiation, this yields for the radiative cooling time
$t_{\rm rad} \sim 10^{12}\,$yrs, which suggests that the protons can
easily maintain their energy for many millions of parsecs without being
seriously effected by synchrotron or inverse-Compton losses, as
expected. For \SgrA, we assume equipartition with $B \sim 10\,$G (Atoyan
\& Dermer 2004b) and $\Gamma_{\rm esc} \sim 7$ (see Table~2, model 5).
The radiative cooling time for the escaping protons is therefore $t_{\rm
rad} \sim 10^8\,$yrs. Hence synchrotron and inverse-Compton losses have
virtually no effect on the energy of the protons in the \SgrA jet.

In addition to synchrotron and inverse-Compton radiation, the protons in
the jet will also lose energy due to Coulomb coupling with the thermal
electrons, which radiate much more efficiently than the protons. The
energy loss rate for this process can be estimated using equation~(4.16)
from Mannheim \& Schlickeiser (1994), which yields
\begeq
\left(dE \over dt\right)\Bigg|_{\rm coul}
= \, 30 \, n_e \, \sig \, m_e c^3
\ ,
\label{eqnew112}
\fineq
where $n_e$ represents the electron number density in the jet.
The associated loss timescale for a proton escaping from the disk
with mean energy $\Gamma_{\rm esc} m_p c^2$ is
\begeq
t_{\rm coul} \equiv {\Gamma_{\rm esc} m_p c^2 \over
~(dE / dt)\big|_{\rm coul}}
= {\Gamma_{\rm esc} \, m_p \over 30 \, n_e \,
\sig \, c \, m_e}
\ .
\label{eqnew113}
\fineq
The electron number density $n_e$ decreases rapidly as the jet expands
from the disk into the external medium. Hence the most conservative
estimate (based on the strongest Coulomb coupling) is obtained by adopting
conditions at the base of the jet, where $n_e$ has its maximum value. To
estimate the electron number density at the base of the outflow, we
begin by calculating the rate at which protons escape from the disk at
the shock location. By using equation~(\ref{Beq8}) to eliminate $A_0$ in
equation~(\ref{eqnew75}), we find that the proton escape rate is given by
\begeq
\Nesc = {4 \pi r_* \lambda_{\rm mag}^2 \, c \, n_*
\over H_*}
\ ,
\label{eqnew114}
\fineq
where $r_*$, $n_*$, $H_*$, and $\lambda_{\rm mag}$ denote the radius,
the proton number density, the vertical half-thickness, and the magnetic
mean free path inside the disk at the shock location, respectively. The
shock is expected to have a width comparable to $\lambda_{\rm mag}$, and
therefore the sum of the upper and lower face areas of the shock annulus
is equal to $4 \pi r_* \lambda_{\rm mag}$. We also note that the flux of
the relativistic protons escaping from the disk into the outflow is
given by $c n_p$, where $n_p$ is the proton number density at the base
of the jet. Combining these relations, we can write the proton escape
rate in terms of $n_p$ using
\begeq
\Nesc = 4 \pi r_* \lambda_{\rm mag} \, c \, n_p
\ .
\label{eqnew115}
\fineq

By equating the two expressions for $\Nesc$ given by
equations~(\ref{eqnew114}) and (\ref{eqnew115}), we find that $n_p$ is
related to $n_*$ via
\begeq
{n_p \over n_*} = {\lambda_{\rm mag} \over H_*} \ < \ 1
\ .
\label{eqnew116}
\fineq
Since the electron-proton jet must be charge neutral, the electron number
density at the base of the jet, $n_e$, is equal to the proton number
density $n_p$, and therefore we obtain
\begeq
n_e = {\lambda_{\rm mag} \over H_*} \ n_*
\ .
\label{eqnew117}
\fineq
Using the relation $\lambda_{\rm mag}/H_*=A_0^{1/2}$ (see
eq.~[\ref{Beq8}]) along with the results for $A_0$ and $n_*$ reported in
Table~1 for M87 (model~2), we obtain $n_e = 0.11\,n_*=2.2 \times 10^3
\,\rm cm^{-3}$. Setting $\Gamma_{\rm esc} \sim 8$, we find that
equation~(\ref{eqnew113}) yields for the electron-proton Coulomb
coupling timescale $t_{\rm coul} \sim 3.5 \times 10^5\,$yrs. Note that
this is an extremely conservative estimate since it is based on
conditions at the bottom of the jet, and therefore it suggests that
Coulomb coupling between the protons and the electrons is insufficient
to seriously degrade the energy of the accelerated ions escaping from
the disk as they propagate out to the radio lobes via the jet. For
\SgrA, we use the model~5 data in Table~2 to obtain $n_e = 0.13\,n_*=5.6
\times 10^4 \,\rm cm^{-3}$. Setting $\Gamma_{\rm esc} \sim 7$
yields for the Coulomb coupling timescale $t_{\rm coul} \sim 1.2 \times
10^4\,$yrs, which implies that the length of the jet can be as large
several thousand parsecs before much energy is drained from the protons,
assuming the material in the jet travels at half the speed of light. We
emphasize that these numerical estimates of the importance of radiative
and Coulomb losses experienced by the relativistic protons are based on
the ``worst-case'' assumption that the conditions at the base of the
outflow prevail throughout the jet. In reality, the jet density will
drop rapidly as the gas expands, and therefore the true values for the
proton energy loss timescales will be much larger than the results
obtained here. This strongly suggests that shock acceleration of the
protons in the disk, as investigated here, is sufficient to power the
observed outflows without requiring any reacceleration in the jets.

\subsection{Radiative Losses from the Disk}

In the ADAF scenario that we have focused on, radiative losses from the
disk are ignored. The self-consistency of this approximation can be
evaluated by computing the free-free emissivity due to the thermal gas
in the disk. The total X-ray luminosity can be estimated by integrating
equation~(5.15b) from Rybicki \& Lightman (1979) over the disk volume to
obtain for pure, fully-ionized hydrogen
\begeq
L_{\rm rad} = \int_{\rs}^\infty 1.4 \times 10^{-27} \,
T_e^{1/2} \, \rho^2 \, m_p^{-2} \, dV \, ,
\label{eqnew118}
\fineq
where $dV = 4 \pi r H dr$ represents the differential (cylindrical)
volume element, and $T_e$ denotes the electron temperature. We can
obtain an upper limit on the X-ray luminosity by assuming that the
electron temperature is equal to the ion temperature. Based on the
detailed disk structures associated with models 2 and 5, we find that
$L_{\rm rad}/\Ljet \sim 10^{-2}$ and $L_{\rm rad}/\Ljet \sim 10^{-5}$,
respectively. However, in an actual ADAF disk, the X-ray luminosity will
of course be substantially smaller than these values because the
electron temperature is roughly three orders of magnitude lower than the
ion temperature. Hence our neglect of radiative losses is completely
justified, as expected for ADAF disks.

\section{CONCLUSION}

In this paper we have demonstrated that particle acceleration at a
standing, isothermal shock in an ADAF accretion disk can energize the
relativistic protons that power the jets emanating from radio-loud
sources containing black holes. The work presented here represents a new
type of synthesis that combines the standard model for a transonic ADAF
flow with a self-consistent treatment of the relativistic particle
transport occurring in the disk. The energy lost from the background
(thermal) gas at the isothermal shock location results in the
acceleration of a small fraction of the background particles to
relativistic energies. One of the major advantages of our coupled,
global model is that it provides a single, coherent explanation for the
disk structure and the formation of the outflow based on the
well-understood concept of first-order Fermi acceleration in shock
waves. The theory employs an exact mathematical approach in order to
solve simultaneously the combined hydrodynamical and particle transport
equations.

The analysis presented here closely parallels the early studies of
cosmic-ray shock acceleration. As in those first investigations (e.g.,
Blandford \& Ostriker 1978), we have employed an idealized model in
which the pressure of the accelerated particles is assumed to be
negligible compared with that of the thermal background gas (the ``test
particle'' approximation). In order to check the self-consistency of
this assumption, we have confirmed that the total pressure is dominated
by the pressure of the background (thermal) gas throughout most of the
disk. However, in the vicinity of the shock the two pressures can become
comparable and this suggests that the dynamical results will change
slightly if the test particle approximation is relaxed. We plan to
consider this question in future work by developing a ``two-fluid''
version of our model that includes the particle pressure, in analogy
with the ``cosmic-ray modified shock'' scenario for cosmic-ray acceleration
(Becker \& Kazanas 2001; Drury \& V\"olk 1981).

We have presented detailed results that confirm that the general
properties of the jets observed in M87 and \SgrA can be understood
within the context of our disk/shock/outflow model. In particular, our
results indicate that the shock acceleration mechanism can produce
relativistic outflows with terminal Lorentz factors and total powers
comparable to those observed in M87 and \SgrA. However, in principle
even higher efficiencies can be achieved by varying the upstream energy
transport rate $\epsilon_-$ which is the fundamental free parameter in
our model. The buildup of the particle pressure in such high-efficiency
situations would require relaxation of the test-particle approximation,
as discussed above. In this paper we have focused on inviscid disks,
which are the simplest to analyze. While the inviscid model provides
useful insight into the importance of shock acceleration in ADAF disks,
this restriction clearly must be lifted in the future, since viscosity
plays a key role in determining the structure of an actual accretion
disk. We are currently developing a self-consistent viscous disk model
in order to explore shock formation and particle acceleration in a more
rigorous context. However, we do not expect the presence of viscosity to
alter any of the basic conclusions reached in this paper because
significant particle acceleration will occur regardless of the
viscosity, provided a shock is present. The existence of shocks in
viscous disks is a controversial issue, but several studies suggest that
shock formation is possible provided the viscosity is relatively low.
In the absence of a consensus regarding the possible presence of shocks
in accretion disks, we believe that it is important to study models
with shocks in order to develop theoretical predictions that can be
tested observationally.

The shock acceleration mechanism analyzed in this paper is effective
only in rather tenuous, hot disks, and therefore we conclude that our
model may help to explain the observational fact that the brightest X-ray
AGNs do not possess strong outflows, whereas the sources with low X-ray
luminosities but high levels of radio emission do. We suggest that the
gas in the luminous X-ray sources is too dense to allow efficient Fermi
acceleration of a relativistic particle population, and therefore in
these systems, the gas simply heats as it crosses the shock. Conversely,
in the tenuous ADAF accretion flows studied here, the relativistic
particles are able to avoid thermalization due to the long collisional
mean free path, resulting in the development of a significant nonthermal
component in the particle distribution which powers the jets and
produces the strong radio emission. We therefore conclude that the
coupled, self-consistent theory for the disk structure and the particle
acceleration investigated here is capable of powering the outflows
observed in many radio-loud systems containing black holes.

The authors are grateful to Dr. Lev Titarchuk for providing a number
of useful comments on the manuscript, and also to the anonymous referee
for several insightful suggestions that significantly improved the paper.

\newpage

\section*{APPENDICES}

\appendix

\section{Treatment of the Vertical Structure}

In principle, the pressure $P$, density $\rho$, diffusion coefficient
$\kappa$, Green's function $\green$, and velocity components $v_r$ and
$v_z$ in the disk all display significant variations in the vertical
($z$) direction. Following Abramowicz \& Chakrabarti (1990), we will
use the first five quantities to represent vertical averages over the
disk structure at radius $r$. However, the vertical variation of the velocity
component $v_z$ must be treated differently. Here, we assume for simplicity
that the vertical expansion is {\it homologous}, and therefore the vertical
velocity variation is given by
\begeq
v_z(r,z) = B(r) \, z \ .
\label{Aeq1}
\fineq
It follows that the vertical velocity at the surface of the disk,
$z=H(r)$, can be written as
\begeq
v_z(r,z)\bigg|_{z=H} = B(r) \, H(r) \ .
\label{Aeq2}
\fineq
In a steady-state situation, we can also express the vertical velocity
at the disk surface using
\begeq
v_z(r,z)\bigg|_{z=H} = v_r {d H \over d r} \ .
\label{Aeq3}
\fineq
By combining the two previous expressions, we find that the function
$B(r)$ is given by
\begeq
B(r) = v_r {d \ln H \over dr} \ .
\label{Aeq4}
\fineq
This result will prove useful when we vertically integrate the transport
equation. Note that in terms of $B(r)$, we can write the divergence of
the flow velocity $\vec v$ in cylindrical coordinates as
\begeq
\vec\nabla \cdot \vec v = {1 \over r} {\partial \over \partial r}
\left(r v_r\right) + {\partial v_z \over \partial z} =
{1 \over r} {\partial \over \partial r}
\left(r v_r\right) + B(r) \ ,
\label{Aeq5}
\fineq
where we have assumed azimuthal symmetry. Application of
equation~(\ref{Aeq4}) now yields
\begeq
\vec\nabla \cdot \vec v = {1 \over H r}
{\partial \over \partial r} \left(r H v_r\right) \ .
\label{Aeq6}
\fineq

The steady-state transport equation expressed in cylindrical
coordinates is (see eq.~[\ref{eqnew60}])
\begeqarray
v_r {\partial \green \over \partial r}
+ v_z {\partial \green \over \partial z}
&=& {1 \over 3} \left[{1 \over r} {\partial \over \partial r}
\left(r v_r\right) + {d v_z \over dz}\right]
E {\partial \green \over \partial E}
+ {1 \over r} {\partial \over \partial r}
\left(r \kappa {\partial \green \over \partial r}\right)
\nonumber \\
&+& {\N0 \, \delta(E-E_0) \, \delta(r-r_*) \over
(4 \pi E_0)^2 \, r_* \, H_*} - A_0 \, c \, \delta(r-r_*) \, \green \ .
\label{Aeq7}
\fineqarray
Operating on equation~(\ref{Aeq7}) with $\int^\infty_0 dz$ and
applying equation~(\ref{Aeq1}) yields, after partially
integrating the term containing $v_z$ on the left-hand side,
\begeqarray
v_r {\partial \over \partial r}(H \green) - H B \green
&=& {1 \over 3} \left[{1 \over r}{d \over d r} \left(r v_r\right)
+ B \right] H E {\partial \green \over \partial E}
+ {1 \over r} {\partial \over \partial r}
\left(r H \kappa {\partial \green \over \partial r}\right)
\nonumber \\
&+& {\N0 \, \delta(E-E_0) \, \delta(r-r_*) \over
(4 \pi E_0)^2 \, r_*} - A_0 \, c \, H_* \,
\delta(r-r_*) \, \green \ ,
\label{Aeq8}
\fineqarray
where the symbols $\green$, $v_r$, and $\kappa$ now refer to vertically
averaged quantities. Using equations~(\ref{Aeq4}), (\ref{Aeq5}),
and (\ref{Aeq6}), we can rewrite the vertically integrated transport
equation as
\begeqarray
H v_r {\partial \green \over \partial r}
&=& {1 \over 3 r} {\partial \over \partial r} \left(
r H v_r \right) E {\partial \green \over \partial E}
+ {1 \over r} {\partial \over \partial r}
\left(r H \kappa {\partial \green \over \partial r}\right)
\nonumber \\
&+& {\N0 \, \delta(E-E_0) \, \delta(r-r_*) \over
(4 \pi E_0)^2 \, r_*} - A_0 \, c \, H_* \,
\delta(r-r_*) \, \green \ .
\label{Aeq9}
\fineqarray

\section{Derivation of the Escape Parameter}

The dimensionless parameter $A_0$ appearing in equation~(\ref{eqnew57})
determines the rate of particle escape through the surface of the disk
due to random walks occurring near the shock location. Since the
particles are accelerated as a consequence of collisions with magnetic
waves, we will assume that the thickness of the shock is comparable to the
magnetic mean free path, $\lambda_{\rm mag}$. In order to estimate $A_0$,
we model the escape of the particles from the disk using the analogy
of ``leakage'' through an opening in a cylindrical pipe with radius equal
to the half-thickness of the disk at the shock location, $H_*$. The length
of the open section of the pipe is set equal to the shock thickness
$\lambda_{\rm mag}$. The particle number density in the open section
is governed by the equation
\begeq
v_x {dn_r \over dx} = - {n_r \over t_{\rm esc}} \ ,
\label{Beq1}
\fineq
where $v_x$, $n_r$, and $t_{\rm esc}$ represent the flow velocity,
the relativistic particle number density, and the average time for the
particles to escape through the open walls of the pipe via diffusion.
Upon integration, the solution to equation (\ref{Beq1}) is given by
\begeq
n_r(x) = n_0 \, \exp\left(-{x \over v_x t_{\rm esc}}\right) \ ,
\label{Beq2}
\fineq
where $n_0$ is the incident number density as the flow encounters the
opening in the pipe, at $x=0$. We can approximate the solution for
$n_r(x)$ by performing a Taylor expansion around $x=0$, which yields
\begeq
n_r(x) \approx n_0 \left(1 - {x \over v_x t_{\rm esc}}\right)\ .
\label{Beq3}
\fineq
The fraction of particles that escape from the pipe can therefore
be estimated by setting $x = \lambda_{\rm mag}$ to obtain
\begeq
f_{\rm esc} = 1 - {n_r \over n_0} = {\lambda_{\rm mag} \over v_x
\, t_{\rm esc}}
\ .
\label{Beq4}
\fineq
In order to make contact with the disk application, we note that
according to equations~(\ref{eqnew74}) and (\ref{eqnew75}), the
fraction of particles that escape as the gas crosses the isothermal
shock is given by
\begeq
f_{\rm esc} = A_0 \, {c \over v_*} \ ,
\label{Beq5}
\fineq
where $v_* \equiv (v_+ + v_-)/2$ is the mean velocity at the shock, and
we have assumed that advection dominates over diffusion. Eliminating
$f_{\rm esc}$ between equations~(\ref{Beq4}) and (\ref{Beq5}), and
setting $v_x = v_*$, we find that
\begeq
A_0 = {\lambda_{\rm mag} \over c \, t_{\rm esc}} \ .
\label{Beq6}
\fineq

Within the context of our one-dimensional model for the particle
transport in the disk, the mean escape time $t_{\rm esc}$ is related to
$\lambda_{\rm mag}$ and the disk half-thickness at the shock $H_*$ via
\begeq
t_{\rm esc} = {H_* \over v_{\rm diff}}
= {H_*^2 \over c \, \lambda_{\rm mag}} \ ,
\label{Beq7}
\fineq
where $v_{\rm diff} = c \lambda_{\rm mag}/H_*$ denotes the vertical
diffusion velocity of the protons in the tangled magnetic field near the
shock, which is valid provided $H/\lambda_{\rm mag} > 1$. Eliminating
$t_{\rm esc}$ between equations~(\ref{Beq6}) and (\ref{Beq7}) then
yields
\begeq
A_0 = \left({\lambda_{\rm mag} \over H_*}\right)^2 \ < \ 1 \ .
\label{Beq8}
\fineq
The diffusion coefficient at the shock is related to the magnetic mean
free path by the standard expression (e.g., Reif 1965)
\begeq
\kappa = {c \, \lambda_{\rm mag} \over 3} \ ,
\label{Beq9}
\fineq
and therefore equation~(\ref{Beq8}) can be rewritten as
\begeq
A_0 = \left({3 \, \kappa_* \over c \, H_*}\right)^2 \ ,
\label{Beq10}
\fineq
where $\kappa_* \equiv (\kappa_- + \kappa_+)/2$ denotes the average of
the upstream and downstream values of $\kappa$ on either side of the shock.

\end{document}